\documentclass[letterpaper,twocolumn,10pt,dvipsnames]{article}
\usepackage{usenix-2020-09}

\usepackage{tikz}
\usepackage{amsmath}
\usepackage{xcolor,soul}
\usepackage{tcolorbox}
\tcbuselibrary{breakable}
\usepackage{MnSymbol}
 \usepackage{multirow}
 \usepackage{enumitem}
 \usepackage{hyperref}

\usepackage{booktabs}
\usepackage{colortbl}
\usepackage[normalem]{ulem}
\useunder{\uline}{\ul}{}

\definecolor{c1}{HTML}{c1a6e3}
\definecolor{c2}{HTML}{a6e3ba}
\definecolor{c3}{HTML}{e0ca79}
\definecolor{c4}{HTML}{e3aba6}
\definecolor{c5}{HTML}{79bce0}

\usepackage{setspace} 

\begin{document}

\date{}

\title{\Large \bf What Was Your Prompt? A Remote Keylogging Attack on AI Assistants}

\author{{Roy Weiss\footnotemark[1], Daniel Ayzenshteyn\footnotemark[1], Guy Amit, Yisroel Mirsky\footnotemark[2]}\\ Ben-Gurion University, Israel\\  Department of Software and Information Systems Engineering  \\
\href{https://offensive-ai-lab.github.io/}{Offensive AI Research Lab} \\ \small \href{https://youtu.be/UfenH7xKO1s}{Link to demo video}
} 

\maketitle

\begin{abstract}
AI assistants are becoming an integral part of society, used for asking advice or help in personal and confidential issues. In this paper, we unveil a novel side-channel that can be used to read encrypted responses from AI Assistants over the web: the token-length side-channel. We found that many vendors, including OpenAI and Microsoft, have this side-channel.

However, inferring the content of a response from a token-length sequence alone proves challenging. This is because tokens are akin to words, and responses can be several sentences long leading to millions of grammatically correct sentences. In this paper, we show how this can be overcome by (1) utilizing the power of a large language model (LLM) to translate these sequences, (2) providing the LLM with inter-sentence context to narrow the search space and (3) performing a known-plaintext attack by fine-tuning the model on the target model's writing style.

Using these methods, we were able to accurately reconstruct 29\% of an AI assistant's responses and successfully infer the topic from 55\% of them. To demonstrate the threat, we performed the attack on OpenAI's ChatGPT-4 and Microsoft's Copilot on both browser and API traffic. 
\end{abstract}
\footnotetext[1]{These authors have equal contribution}
\footnotetext[2]{Corresponding author}

\section{Introduction}

The proliferation of Large Language Models (LLMs) and Chat-based AI services, such as ChatGPT, marks a significant evolution in the digital landscape. These technologies have not only captured the imagination of the public but have also become integral to various aspects of society. Their utility spans from answering simple queries to assisting in complex decision-making processes, highlighting their importance and the trust placed in them by users worldwide.

As the use of AI assistants becomes increasingly commonplace, so too does the sensitivity of the information shared with them. Users frequently turn to these assistants for discussions on personal matters, such as health and hygiene, or for help with confidential tasks, including editing sensitive emails or seeking business advice. This trend underscores the critical need for robust security measures to protect the privacy of these interactions~\cite{addington2023chatgpt, weidinger2022taxonomy}.

\begin{figure}[t]
    \centering
    \includegraphics[width=\columnwidth]{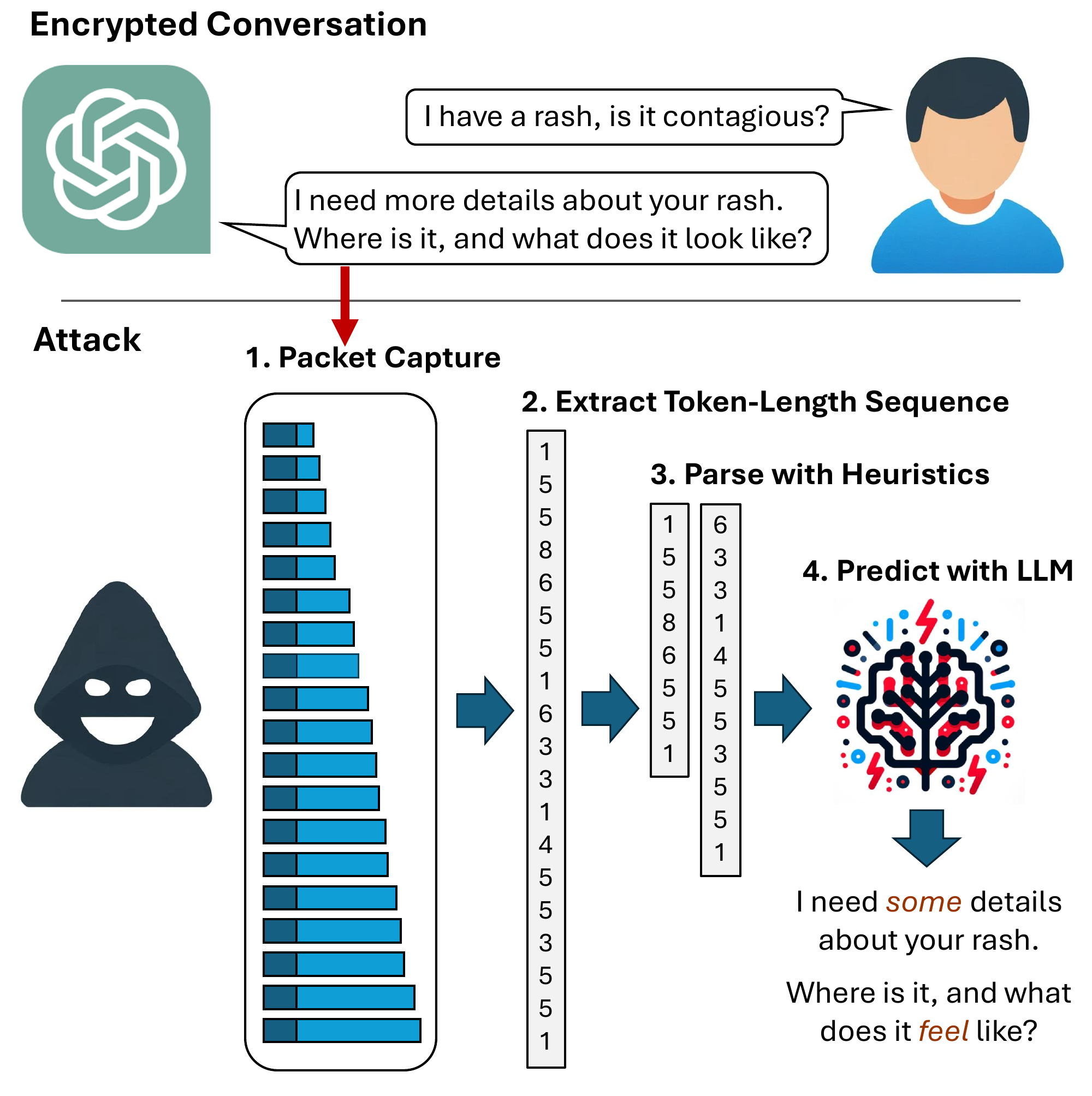}
    \caption{Overview of the attack. A packet capture of an AI assistant's real-time response reveals a token-sequence side-channel. The side-channel is parsed to find text segments which are then reconstructed using sentence-level context and knowledge of the target LLM's writing style.}
    \label{fig:attack}
    \vspace{-1em}
\end{figure}

However, our research has uncovered a significant vulnerability in the way LLM services, including the popular ChatGPT-4~\cite{achiam2023gpt}, handle data transmission. LLMs generate and send responses as a series of tokens (akin to words), with each token transmitted from the server to the user as it is generated. While this process is encrypted, the sequential token transmission exposes a new side-channel: the token-length side-channel. Despite encryption, the size of the packets can reveal the length of the tokens, potentially allowing attackers on the network to infer sensitive and confidential information shared in private AI assistant conversations.  The challenge for attackers in exploiting the token-length side-channel lies in the inherent complexity of accurately inferring text from a sequence of token-lengths. This difficulty is primarily due to the fact that the tokens from a single sentence can correspond to a multitude of grammatically correct sentences. Moreover, this task becomes exponentially more challenging when the goal is to decipher entire paragraphs, vastly increasing the potential combinations and interpretations. 

Previous studies on remote keyloggers have leveraged additional side-channels, such as keystroke timings, to reduce entropy and facilitate the inference of typed information. However, such approaches are not applicable in our setting because LLMs generate tokens for whole words at a time and do not leak character-level information. This presents a unique challenge to traditional side-channel analysis. A complete discussion of related works can be found in section \ref{sec:relworks}.

To overcome this challenge, we propose a \textit{token inference attack} that is extremely effective at deciphering responses in encrypted traffic. The approach is to train a state-of-the-art LLM to translate a token-length sequences back into legible sentences. Furthermore, by providing the LLM with the context of previously inferred sentences, the LLM can narrow down the possible sentences further, thereby reducing the entropy involved in inferring entire paragraphs. Finally, we show how an adversary can exploit the predictable response style and phrase repetition of LLMs like ChatGPT to refine the model's accuracy even further. This is achieved by training the attack model with sample chats from the target AI assistant, effectively creating a known-plaintext attack scenario that enhances the model's ability to infer the sequence of tokens. These sample chats can be readily obtained from public repositories (e.g., \cite{ding2023enhancing}) or by accessing the AI assistant service directly as a paying user.

Our investigation into the network traffic of several prominent AI assistant services uncovered this vulnerability across multiple platforms, including Microsoft Bing AI (Copilot) and OpenAI's ChatGPT-4. We conducted a thorough evaluation of our inference attack on GPT-4 and validated the attack by successfully deciphering responses from four different services from OpenAI and Microsoft.  

In summary, the contributions of our paper are as follows:
\begin{itemize}
    \item \textbf{Novel side-channel}: We identify a novel side-channel inherent in all LLM models, affecting any LLM-based service that sends responses in real-time. 
    \item \textbf{Token Extraction Method}: We provide framework for extracting token-length sequences from encrypted LLM response traffic, and identifying text segments (e.g., sentences) within the sequences.
    \item \textbf{Single Sentence token inference Attack}: We propose the first ever token inference attack. We train an LLM to translate token-length sequences into plaintext sentences. To the best of our knowledge, this is the first work that uses generative AI to \textit{perform} a side-channel attack.
    \item \textbf{Multi-sentence token inference Attack}: We introduce a technique for inferring entire paragraphs by considering inter-sentence context. By doing so, we are able to significantly narrow the scope of possible sentences and enhance the accuracy of the inferred information.
    \item \textbf{Known-Plaintext Attack on LLMs}: We expose a novel attack vector which can be used to improve token inference attacks. By collecting and analyzing example responses from the target LLM (i.e., AI assistant) we can exploit the predictable style of LLMs, and tendency of LLM's to repeat training data, to better infer plaintexts.
    \item \textbf{Exposure \& Demonstration} We identify several major vendors vulnerable to this attack and demonstrate the significance of the attack by demonstrating it on encrypted network traffic from OpenAI's ChatGPT-4 and Microsoft's Copilot.
\end{itemize}

This paper not only sheds light on a critical security flaw in current AI assistant services but also offers a comprehensive framework for understanding and mitigating the risks associated with the token-length side-channel.

\raggedbottom
\section{Background}
In this section, we outline the fundamental concepts needed to understand the token-length side-channel. We start by defining tokens and the tokenization process. Next, we briefly describe how Large Language Models (LLMs) use and generate token sequences. Finally, we examine the deployment of LLMs in online AI assistants and chatbots, highlighting how the streaming of tokens can expose their lengths to potential eavesdroppers.
\subsection{Tokens \& Tokenizers}

In Natural Language Processing (NLP), a token is the smallest unit of text that carries meaning. The set of all tokens $K$ is predetermined based on the content being processed. When a sentence is tokenized, it is divided into a series of tokens represented as \( S = (k_1, k_2, ..., k_n) \), where \( S \) is the entire sentence, and $k_i \in K$ is an individual token.

For instance, consider the sentence ``\textit{I have an itchy rash}.'' The tokenization of this sentence could be represented as \( S = (k_1, k_2, k_3, k_4, k_5) \), where the tokens are $k_1 =$ \textit{I}, $k_2 =$ \textit{have}, $k_3 =$ \textit{an}, $k_4 =$ \textit{itchy}, and $k_5 =$ \textit{rash}.
In some cases, particularly with complex words, tokenization might result in a word being divided into multiple tokens. For example, the word ``\textit{tokenize}'' could be tokenized as \( S = (k_1, k_2) \) with $k_1 =$ \textit{token} and $k_2 =$ \textit{ize}. 
Furthermore, spaces and punctuation in tokenization are handled distinctively. Spaces are often included in the token, while punctuation like commas and periods are typically separate tokens. For example, consider the text ``\textit{Oh no! I'm sorry to hear that. Try applying some cream}.'' The tokenizer of GPT-3.5 and 4 would tokenize it as

\sethlcolor{c1}\hl{Oh}\sethlcolor{c2}\hl{ no}\sethlcolor{c3}\hl{!}\sethlcolor{c4}\hl{ I}\sethlcolor{c5}\hl{'m}\sethlcolor{c1}\hl{ sorry}\sethlcolor{c2}\hl{ to}\sethlcolor{c3}\hl{ hear}\sethlcolor{c4}\hl{ that}\sethlcolor{c5}\hl{.}\sethlcolor{c1}\hl{ Try}\sethlcolor{c2}\hl{ applying}\sethlcolor{c3}\hl{ some}\sethlcolor{c4}\hl{ cream}\sethlcolor{c5}\hl{.}

\noindent and the tokenizer of LLAMA-1 and 2 would tokenize it as

\sethlcolor{c1}\hl{Oh}\sethlcolor{c2}\hl{ no}\sethlcolor{c3}\hl{!}\sethlcolor{c4}\hl{ I}\sethlcolor{c5}\hl{'}\sethlcolor{c2}\hl{m}\sethlcolor{c1}\hl{ sorry}\sethlcolor{c2}\hl{ to}\sethlcolor{c3}\hl{ hear}\sethlcolor{c4}\hl{ that}\sethlcolor{c5}\hl{.}\sethlcolor{c1}\hl{ Try}\sethlcolor{c2}\hl{ applying}\sethlcolor{c3}\hl{ some}\sethlcolor{c4}\hl{ cre}\sethlcolor{c2}\hl{am}\sethlcolor{c5}\hl{.}

\noindent Note how the apostrophe and the letter following it can sometimes form a separate token, and that spaces are added as prefixes to words. 
Tokenizers used by different LLM models are fundamentally similar as they all follow the principle of breaking down text into manageable units. Also, it is important to note that major vendors do not keep their tokenizers secret since they are an important part of their API services.

In summary, although tokens are akin to words, they do not have a one-to-one mapping. They also include spacing and reserve tokens for punctuation.

\subsection{LLMs in AI Assistants}
AI chatbots, like ChatGPT, are sophisticated LLMs designed for engaging in human language interactions. Key to their functionality are prompts and responses:

\begin{description}
    \item[Prompt (\( P \)):] A prompt is the user's input, typically a question or statement, initiating interaction with the LLM. It is represented as a token sequence \( P = [p_1, p_2, ..., p_m] \) for $p_i \in K$.
    \item[Response (\( R \)):] In reply to the prompt, the LLM generates a response, also a sequence of tokens, denoted as \( R = [r_1, r_2, ..., r_n] \) for $r_i \in K$.
\end{description}

Chat-based LLMs manage conversations through these alternating prompts and responses, maintaining context to ensure relevance and coherence. The model tracks the dialogue's history, allowing it to contextualize each response within the ongoing conversation.

LLMs use tokens both in training and execution. They are trained on vast datasets of tokenized text (e.g. The Pile~\cite{gao2020pile}) to learn the probability of a token following a given sequence, which enables them to predict the next token in a response.

During execution, the LLM generates response tokens \textbf{sequentially}. Starting from the prompt, it predicts each subsequent token \( r_i \) based on both the prompt and the preceding response tokens \( [r_1, r_2, ..., r_{i-1}] \). This method, \( p(r_i | P, r_1, r_2, ..., r_{i-1}) \), allows the LLM to produce contextually relevant and coherent responses, considering the entire conversational history.

\subsection{The Deployment of AI Assistants}\label{subsec:deployment}
AI assistants are typically deployed in cloud-based environments. This setup allows for scalable and efficient access to the computational resources required to run these sophisticated models. A user session with an AI assistant generally follows a straightforward process:

\begin{enumerate}
    \item \textbf{Connection}: The user connects to a server hosted in the cloud via a web app in a browser or via an API (e.g., using a 3rd-party app). The user starts or resumes a chat session (conversation) to set the context of the prompts.
    \item \textbf{Prompting}: The user submits a prompt $P$ (a query or statement) and it is transmitted to the server as a single message. The server forwards the prompt to an instance of the LLM model for processing.
    \item \textbf{Response Generation}: The LLM generates a response $R$ to the prompt and the response tokens are sent back to the user \textbf{sequentially} and \textbf{in real time} for visualizing the response as it's created. This operational approach enhances user experience by allowing users to see the AI's responses form in real-time, ensuring a dynamic and engaging conversation. This is especially important given that state-of-the-art LLMs are slow due to their complexity. 
\end{enumerate}

We observed that most vendors use either the QUIC protocol over UDP or web sockets over TCP to transmit responses. As of the time of writing, these vendors do not pad, compress, or encode the traffic before it is encrypted. For more details, see Table \ref{tab:vendors}.

\begin{table}[]
\resizebox{\columnwidth}{!}{%
\setlength\tabcolsep{1.5pt} 

\begin{tabular}{ccc|cccc|cccccc|cc|ccc|}
                                                    &                                                      &  \cellcolor[HTML]{C0C0C0}                            & \multicolumn{4}{c|}{LLM Model}                & \multicolumn{6}{c|}{Protocol}                                         & \multicolumn{2}{c|}{Extraction}           & \multicolumn{3}{c|}{Tokens}       \\
{Vendor}                            & {Service}                            & {\cellcolor[HTML]{C0C0C0}\rotatebox{90}{Vulnerable}} & \rotatebox{90}{GPT-4}     & \rotatebox{90}{Claude}    & \rotatebox{90}{Palm2}     & \rotatebox{90}{GPT-3.5}   & \rotatebox{90}{HTTP2}     & \rotatebox{90}{Websocket} & \rotatebox{90}{QUIC}      & \rotatebox{90}{TLS}       & \rotatebox{90}{TCP}       & \rotatebox{90}{UDP}       & \rotatebox{90}{$|m_i|-|m_{i-1}|$} & \rotatebox{90}{$|\text{packet}_i|-h$} & \rotatebox{90}{Buffered}  & \rotatebox{90}{Paired}    & \rotatebox{90}{Single}    \\ \hline
\multicolumn{1}{c|}{OpenAI}                         & \multicolumn{1}{c|}{ChatGPT}                         &\cellcolor[HTML]{C0C0C0} $\bullet$                    & $\bullet$ &           &           & $\bullet$ &           &           & $\bullet$ & $\bullet$ &           & $\bullet$ & $\bullet$         &                       & $\bullet$ &           &           \\
\multicolumn{1}{c|}{OpenAI}                         & \multicolumn{1}{c|}{Marketplace}                 & \cellcolor[HTML]{C0C0C0}$\bullet$                    & $\bullet$ &           &           &           &           &           & $\bullet$ & $\bullet$ &           & $\bullet$ & $\bullet$         &                       & $\bullet$ &           &           \\
\multicolumn{1}{c|}{OpenAI}                         & \multicolumn{1}{c|}{API}                             & \cellcolor[HTML]{C0C0C0}$\bullet$                    & $\bullet$ &           &           &           &           &           &           & $\bullet$ & $\bullet$ &           &                   & $\bullet$             &           & $\bullet$ &           \\
\multicolumn{1}{c|}{Microsoft}                      & \multicolumn{1}{c|}{Copilot}                         & \cellcolor[HTML]{C0C0C0}$\bullet$                    & $\bullet$ &           &           &           &           & $\bullet$ &           & $\bullet$ & $\bullet$ &           & $\bullet$         &                       & $\bullet$ &           &           \\
\multicolumn{1}{c|}{Writesonic}                     & \multicolumn{1}{c|}{Chatsonic}                       & \cellcolor[HTML]{C0C0C0}$\bullet$                    & $\bullet$ &           &           &           & $\bullet$ &           &           & $\bullet$ & $\bullet$ &           &                   & $\bullet$             & $\bullet$ &           &           \\
\multicolumn{1}{c|}{Anthropic}                      & \multicolumn{1}{c|}{Claude AI}                       & \cellcolor[HTML]{C0C0C0}$\bullet$                    &           & $\bullet$ &           &           & $\bullet$ &           &           & $\bullet$ & $\bullet$ &           &                   & $\bullet$             & $\bullet$ &           &           \\
\multicolumn{1}{c|}{Notion}                         & \multicolumn{1}{c|}{AI Copilot}               & \cellcolor[HTML]{C0C0C0}$\bullet$                    &           &           &           &           & $\bullet$ &           &           & $\bullet$ & $\bullet$ &           &                   & $\bullet$             &           & $\bullet$ &           \\
\multicolumn{1}{c|}{ClickUp}                        & \multicolumn{1}{c|}{AI Brain}                & \cellcolor[HTML]{C0C0C0}$\bullet$                    &           &           &           &           & $\bullet$ &           &           & $\bullet$ & $\bullet$ &           &                   & $\bullet$             &           & $\bullet$ &           \\
\multicolumn{1}{c|}{TextCortex}                     & \multicolumn{1}{c|}{TextCortex AI}           & \cellcolor[HTML]{C0C0C0}$\bullet$                    &           &           &           &           & $\bullet$ &           &           & $\bullet$ & $\bullet$ &           &                   & $\bullet$             &           & $\bullet$ &           \\
\multicolumn{1}{c|}{CKSoruce}                       & \multicolumn{1}{c|}{CKEditor}                        & \cellcolor[HTML]{C0C0C0}$\bullet$                    &           &           &           &           & $\bullet$ &           &           & $\bullet$ & $\bullet$ &           &                   & $\bullet$             &           & $\bullet$ &           \\
\multicolumn{1}{c|}{Cloudflare}                       & \multicolumn{1}{c|}{Workers AI}                        & \cellcolor[HTML]{C0C0C0}$\bullet$                    &           &           &           &           &  &           &           & $\bullet$ & $\bullet$ &           &                   & $\bullet$             &           & & $\bullet$           \\

\multicolumn{1}{c|}{Quora}                          & \multicolumn{1}{c|}{Poe}                             & \cellcolor[HTML]{C0C0C0}$\circ$                      & $\bullet$ & $\bullet$ &           & $\bullet$ &           & $\bullet$ &           & $\bullet$ & $\bullet$ &           & $\bullet$         &                       &           &           & $\bullet$ \\
\multicolumn{1}{c|}{Perplexity AI}                  & \multicolumn{1}{c|}{Perplexity AI}                   & \cellcolor[HTML]{C0C0C0}$\circ$                      & $\bullet$ & $\bullet$ &           & $\bullet$ &           & $\bullet$ &           & $\bullet$ & $\bullet$ &           & $\bullet$         &                       &           &           & $\bullet$ \\
\multicolumn{1}{c|}{CopyAI}                         & \multicolumn{1}{c|}{CopyAI}                          & \cellcolor[HTML]{C0C0C0}$\circ$                      &           &           &           &           &           & $\bullet$ &           & $\bullet$ & $\bullet$ &           & $\bullet$         &                       &           &           & $\bullet$ \\
\rowcolor[HTML]{F2F2F2} 
\multicolumn{1}{c|}{\cellcolor[HTML]{F2F2F2}Google} & \multicolumn{1}{c|}{\cellcolor[HTML]{F2F2F2}Bard}    &  \cellcolor[HTML]{C0C0C0}                            &           &           & $\bullet$ &           & $\bullet$ &           &           & $\bullet$ & $\bullet$ &           &                   &                       &           &           &           \\
\rowcolor[HTML]{F2F2F2} 
\multicolumn{1}{c|}{\cellcolor[HTML]{F2F2F2}Github} & \multicolumn{1}{c|}{\cellcolor[HTML]{F2F2F2}Copilot} &  \cellcolor[HTML]{C0C0C0}                            & $\bullet$ &           &           &           &           &           &           & $\bullet$ & $\bullet$ &           &                   &                       &           &           &          
\end{tabular}%
}
\caption{AI Assistants and their vulnerability to the side channel attack, as of the time of writing. ($\bullet=$yes $\circ=$maybe) }
\label{tab:vendors}
\end{table}

\section{Side-channel Definition \& Attack Model}

\subsection{Token-length side-channel}\label{subsec:sidechannel}

In a real-time communication setting, AI services transmit the next token \( r_i \) immediately after it is generated. Our observations of several AI assistant services (referenced in Table \ref{tab:vendors}) indicate that the token \( r_i \) is sent either as an individual message or as part of a cumulative message (e.g., \( [r_1, r_2, ..., r_i] \)). Crucially, in both scenarios, the packet's payload length is directly correlated to the number of characters in \( r_i \). In the case of cumulative messages, the length of each token can be inferred by calculating the difference in payload length between successive packets. Consequently, for each response message, it is possible to discern the lengths of every single token, even when the traffic is encrypted.

Let the token-length sequence for a response be denoted as \( T = [t_1, t_2, ..., t_n] \), where \( t_i \) represents the length of the token \( r_i \). The relationship between the token \( r_i \) and its length \( t_i \) can be expressed as \( t_i = |r_i| \), the absolute number of characters in \( r_i \).

This token-length sequence \( L \) can be exploited to infer the original tokens, thereby breaching the privacy of the conversation by revealing every AI response. These responses can also be used to deduce the prompts themselves, either indirectly through context or directly in cases where the AI repeats the question before proceeding.

\subsection{Attack Model}\label{subsec:attackmodel}
In our scenario, we have three entities: Bob, the user; Alice, the AI assistant; and Eve, the attacker. Bob interacts with Alice for various tasks such as seeking personal advice, looking up facts, or editing documents. Alice, the AI assistant, responds to Bob’s prompts over an encrypted communication channel. We assume that all plaintext messages exchanged are in English and that no additional padding is added to these messages, as is common practice with several major vendors (see Table \ref{tab:vendors}).

Eve, positioned as the adversary in this model, aims to read the encrypted responses sent by Alice. She is capable of observing the encrypted network packets, either within the Local Area Network (LAN) of Bob or somewhere in the internet infrastructure between Alice and Bob. By monitoring these packets, Eve intends to extract the token-length sequence \( T \) from each response and use it to infer the original plaintext \( R \). With \( R \), Eve can access not only private and personal information about Bob but also potentially sensitive data related to the company Bob works for. For instance, this could occur if Bob asks Alice for assistance in editing a work-related email.

Although we don't require it, we assume that Eve has access to publicly available datasets of example prompts and responses from the target AI service, or that she can register as a free or paid user to create her own dataset.

\subsection{Problem Statement}
The fundamental challenge for the attacker in this scenario is to accurately reconstruct the original response \( R \) from the observed token-length sequence \( T \). 

\noindent This can be formally stated as follows:

\textit{Given a sequence of token-lengths \( T = [t_1, t_2, ..., t_n] \) extracted from encrypted traffic, infer the original token sequence \( R = [r_1, r_2, ..., r_n] \) from \( T \), where \( t_i = |r_i| \) represents the length of token \( r_i \).}

Solving this problem is non-trivial due to the absence of additional character-level information that can help narrow down the options for each word. This lack of information significantly increases the number of possible grammatically correct sentences for even a single sentence structure.

For example, consider the sentence "She has a \_\_\_ and a \_\_\_," with blanks to be filled by 4-letter and 5-letter nouns respectively. There are at least 880 nouns with 4 letters and 905 nouns with 5 letters, so the total number of grammatically correct solutions for the sentence is \( 880 \times 905 = 795,600 \). This example shows that even with just two unknown tokens, brute-forcing all possible \textit{correct} sentences that can be mapped to \( T \) leads to too many results for the attacker to consider. This issue is compounded when considering entire paragraphs.

However, there are at least two sources that can be used to reduce the entropy of this task: long-distance language structure across sentences and paragraphs, and style-specific language structure exhibited by AI assistants. Therefore, we address this challenge by combining three strategies:

\begin{figure*}
    \centering
    \includegraphics[width=\textwidth]{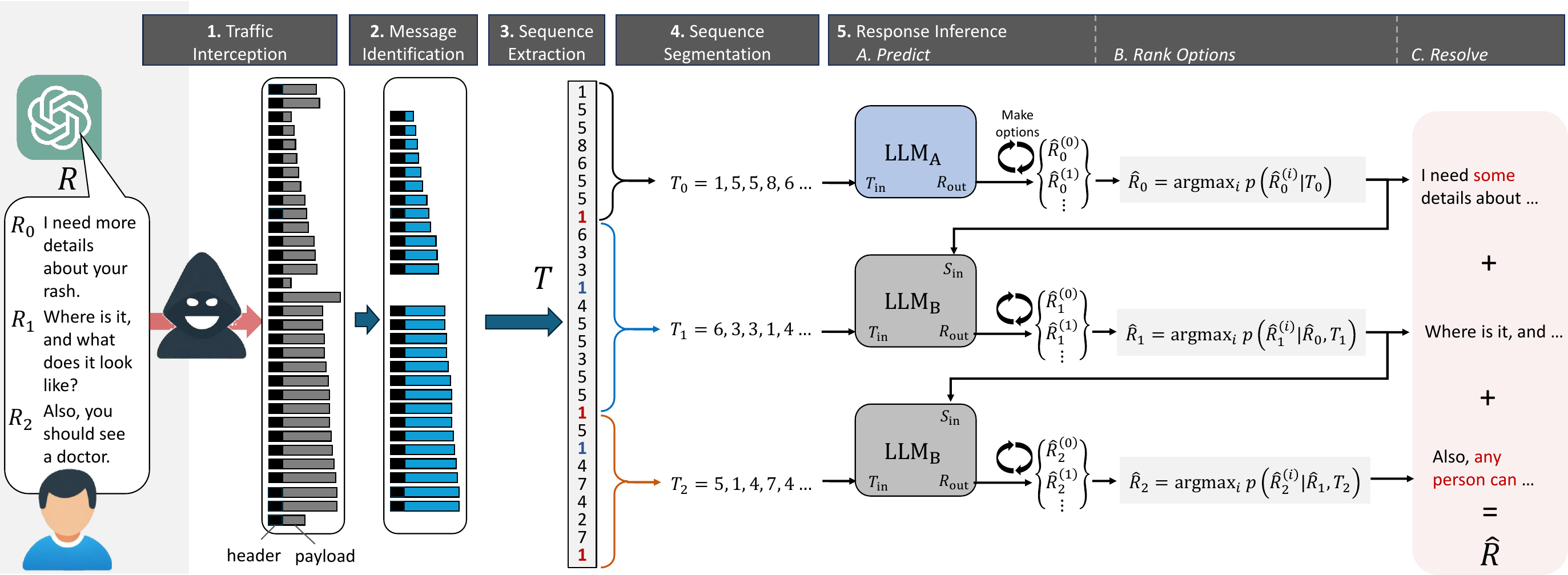}
    \vspace{-1.5em}
    \caption{An overview of the attack framework: (1) Encrypted traffic is intercepted and then (2) the start of the response is identified. Then (3) the token-length sequence $T$ is extracted and (4) a heuristic is used to partition $T$ into ordered segments ($T_0, T_1, ...$). Finally, (5) each segment is used to infer the text of the response.     
    This is done by (A) using two specialized LLMs to predict each segment sequentially based on prior outputs, (B) generating multiple options for each segment and selecting the best (most confident) result, and (C) resolving the predicted response $\hat{R}$ by concatenating the best segments together.}

    \label{fig:framework}
\end{figure*}

\begin{enumerate}
    \item \textbf{Inference with Modern LLMs}: We harness the capabilities of state-of-the-art language models. By fine-tuning a pre-trained LLM for our task, we can capitalize on the common language structures and patterns prevalent in the English language to reduce sentence entropy. Unlike previous works that relied on Markovian models for exploiting similar side-channels (e.g., \cite{monaco2019you}), LLMs are more adept at this task due to their proficiency in considering long-distance relationships between tokens~\cite{vaswani2017attention}. Additionally, we employ self-supervised learning for fine-tuning, which greatly simplifies dataset curation by eliminating the need for manual labeling.
    
    \item \textbf{Forward Context}: Inferring the content of a token sequence is much easier if we know what the previous response was. For example, if $R_{i-1}$ is a sentence about `\textit{how to apply itch cream},' it's likely that $R_i$ will pertain to the application of itch cream. Formally, \( p(R_i | T_i, \hat{R}_{i-1}) \) where $\hat{R}_{i-1}$ is the inferred response from $T_{i-1}$. Therefore, by providing the adversary's LLM with $\hat{R}_{i-1}$ we can greatly reduce paragraph entropy.

Solving for $T$ is hard. However, if $T$ is part of a long sequence (i.e., a sentence in a paragraph) then the context of the prior resolved sentence can help us infer the words in the current sentence better. For example, if the last sentence was about `how to apply itch cream,' then the next sentence may be about `how often to use it.' This context can help an LLM model reduce paragraphs even further. 
    
    \item \textbf{Known-plaintext Attack}: We observed that LLMs used in AI assistant services exhibit distinct writing styles and sometimes repeat phrases from their training data, a notion echoed by other researchers as well~\cite{carlini2022quantifying, nasr2023scalable}. Recognizing this characteristic enables us to conduct an attack similar to a \textit{known-plaintext attack}. The method involves compiling a dataset of responses from the target LLM using public datasets or via sending prompts as a paid user. The dataset can then be used to further fine-tune the inference model. As a result, the inference model is able to reduce entropy significantly, and sometimes even predict the response $R$ from $T$ perfectly, word for word.
\end{enumerate}

\section{Token Inference Attack}
\label{sec:token_length_inference_attack}

In this section, we describe how we implement the \textit{token inference attack} that exploits the \textit{token-length side-channel} exhibited by major vendors.  
Our attack is structured into a series of steps, as depicted in Fig. \ref{fig:framework}:

\begin{enumerate}
    \item  \textbf{Traffic Interception}: The first step involves intercepting the encrypted traffic between the user and the AI assistant. This can be done by eavesdropping on the traffic sent through public networks or by malicious actors within an internet service provider (ISP).
    \item \textbf{Message Identification}: Once the traffic is intercepted, the next step is to extract the message sizes. A message $m$ is a communication  that contains the latest token and other metadata. To do this, we must first identify the first message packet. This involves (1) removing all packets that do not contain messages and (2) combining packets that were split because they were too long. The result is a sequence of message sizes.
    \item \textbf{Sequence Extraction}: With the sequence of message sizes, the token-length sequence \( T \) is extracted by observing the change of the stream's message sizes over time. Depending on the server's approach to token transmission — whether it includes all preceding tokens with each new token or not — two distinct strategies can be employed to extract the token-length sequence \( T \).
    \item \textbf{Sequence Segmentation}: The extracted token-length sequence \( T \) is then partitioned into ordered segments \( T_0, T_1, ...\ \), where $T_i$ roughly corresponds to a sentence. This is accomplished by using a heuristic that exploits the tokenizer's behavior.
    \item \textbf{Response Inference}: The sequence of segments is then passed to a model consisting of two LLMs ($\text{LLM}_\text{A}$ and $\text{LLM}_\text{B}$) which are used to infer the text of $R$. $\text{LLM}_\text{A}$ is designed to reconstruct the first segment from $T_0$ and $\text{LLM}_\text{B}$ is designed to reconstruct the subsequent segments from $T_1$, $T_2$, and so on using the inferred text of the preceding segment as context. We employ two LLMs because the initial sentence of an AI assistant's response typically follows a unique distribution. By tailoring a dedicated model specifically for these first sentences, we enhance the accuracy at the paragraph's outset, which in turn significantly improves the inference quality for all subsequent sentences within the paragraph.
    
   Given the stochastic nature of LLM outputs, for each \(T_i\), we generate multiple outputs and select the most probable one as the predicted response segment \(\hat{R}_i\). These segments, \(\hat{R}_0, \hat{R}_1, ...\), are then concatenated to construct the complete inferred response \(\hat{R}\).
\end{enumerate}

\noindent We will now provide more details on each of these steps.

\subsection{Traffic Interception}
The initial step of the attack involves eavesdropping on encrypted communications, as detailed in Section \ref{subsec:attackmodel}. The adversary can be positioned within the same LAN, connected to the same WiFi, or anywhere on the internet, provided they can observe the packets. Moreover, there exist techniques that allow traffic to be rerouted through the adversary for observation, even if they are not directly on the communication path \cite{conti2016survey}, enhancing the feasibility of eavesdropping. 

To perform the attack, we must find the response. To accomplish this, we filter the traffic based on (1) the server's known IP addresses and (2) the protocol used by the vendor. For example, to target ChatGPT, we search for UDP traffic carrying QUIC, and for Bing Copilot we search for TCP traffic carrying TLS.

\subsection{Message Identification}

\begin{figure}
    \centering
    \includegraphics[width=\columnwidth]{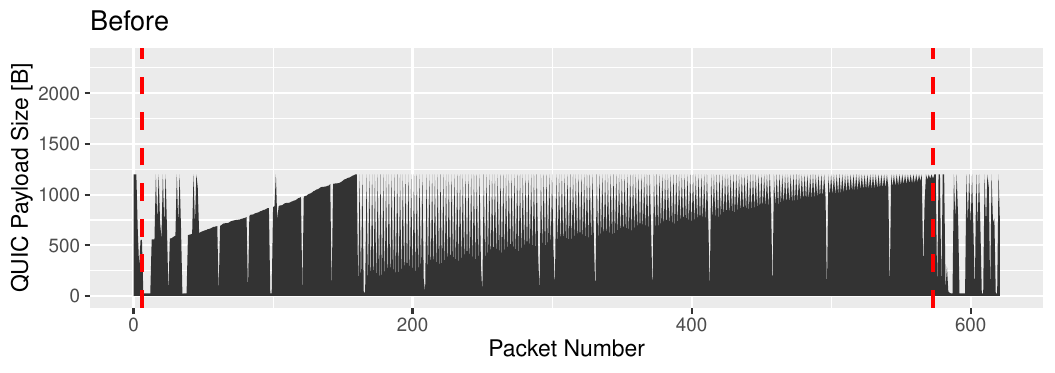}
    \includegraphics[width=\columnwidth]{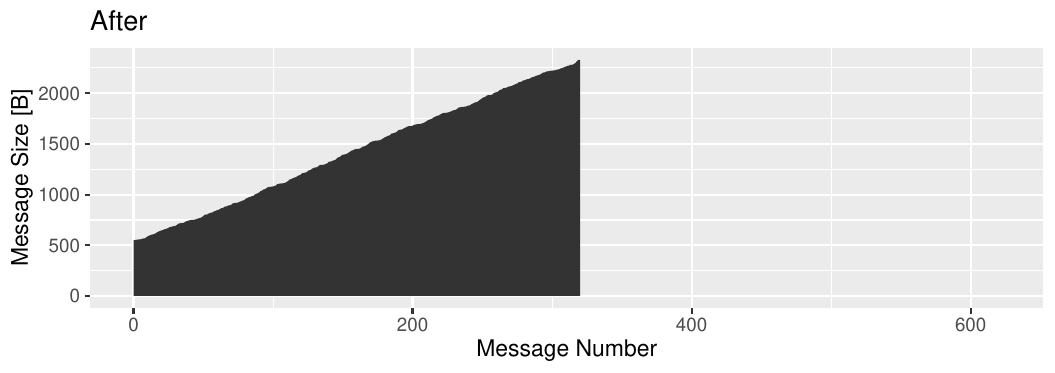}
    \vspace{-2em}
    \caption{An example showing the trends in the encrypted traffic traffic before and after performing message identification. This example is taken from a response sent from OpenAI's ChatGPT-4 web app. Red bars indicate the start and end of the messages.}
    \label{fig:packetsizeIncrease}
    \vspace{-1em}
\end{figure}

As mentioned earlier in section \ref{subsec:sidechannel}, we observed that when the server wants to send the client token $r_i$, it sends a single message $m_i$ containing all subsequent tokens \( r_1, r_2, ..., r_i \) and some fixed length metadata about the session. In this step, our objective is to extract the message sizes from traffic.

To find the packet containing $m_0$, we use a naive yet effective approach: 
Since a given $m_i$ can typically fit into one packet, the sequence of $|R|$ packets will have payload sizes that are \textbf{strictly increasing in length} (see Fig. \ref{fig:packetsizeIncrease}). Therefore, to find the first packet (containing $m_0$), we simply look for this trend and then backtrack to the starting point. Then, using the lengths of these packets, we can infer that $|m_0|$ is equal to that packet's size (excluding the upper layer's header length). 
We note that some vendors use higher-level protocols that send bursts of control packets intermittently. We remove these by expecting continuity in the increasing payload lengths. 

For vendors using QUIC, messages are split over multiple packets when they become too big. For example, for ChatGPT-4's web browser app, the maximum QUIC payload size is 1200 Bytes. When the encapsulated HTTP3 packet exceeds this, it spills over into multiple QUIC packets, each with its own 28 Byte HTTP3 header.
This results in a `saw tooth' pattern in the QUIC payload sizes when $|m|>1200-28$ Bytes. However, this pattern is deterministic: if $|m| > 1172$ Bytes, then there will be $\bigl\lfloor \frac{|m|}{1172}  \bigl\rfloor$ subsequent packets with the maximum payload size and one subsequent packet with a payload size of $| m | \bmod (1172) + 28$ bytes.


\subsection{Sequence Extraction}
Now that we have extracted a sequence of message sizes, we can derive the token-length sequence as follows:
\begin{equation}
    T = \Bigl\{t_i \Bigl\vert \hspace{0.3em} |m_i| - |m_{i-1}| \Bigl\}
\end{equation}
This method is feasible because, as outlined in section \ref{subsec:deployment}, vendors typically do not apply padding, compression, or encoding to tokens prior to encryption, enabling the straightforward extraction of \(T\). It's important to highlight that while all major vendors currently transmit tokens in this manner, future services might opt to send each token independently without including the previous tokens. In such scenarios, one can deduce the overhead and calculate \(T\) as \(T= \{t_i \bigl\vert  |\text{packet}_{i}|-h\}\), where $h$ is the fixed-length message metadata. The value of $h$ can be determined either by connecting as a legitimate user and inspecting the packets or heuristically by identifying the smallest packet likely containing a single-character token.

We note that even if a vendor encodes or compresses the data, it may not completely mitigate the side-channel, as shown in \cite{monaco2019you}.

\subsection{Sequence Segmentation}\label{subsec:sequence_segmentation}
The token-sequence \(T\) encapsulates one or more paragraphs of text. Before we can use it with our model, we must segment it into meaningful chunks. Fortunately, it is possible to identify punctuation marks within \(T\). This is due to two key observations: (1) tokenizers are designed not to include a space character with a punctuation mark, and (2) the smallest word, such as `a', always comes with a leading space, making it at least two characters long. Consequently, tokens of length 1 are almost certainly punctuation marks.

Naturally, for our model to effectively infer text, we aim to segment \(T\) into units that closely resemble complete sentences. However, distinguishing between different types of punctuation marks (periods, commas, etc.) is not feasible directly. Thus, we employ a heuristic to approximate sentence boundaries.

\textbf{Segmentation Process.} The process begins by splitting \(T\) at every instance where \(t_i=1\), which likely indicates a punctuation mark. If a segment contains fewer than 10 tokens ($\sim$7 words), it is merged with the following segment. This step is iterated until we compile a list of token-length sequences (\(T_0, T_1, ...\)), each approximating a sentence or meaningful textual segment. There are some edge cases that can also be considered to refine the results. For example, responses that contain an enumerated list always have the pattern ``:\textbackslash n\textbackslash n1.'' which is the sequence (3,1,1) since the colon and newlines are joined. Therefore, we add this sub-string to the start of the next segment so that each item will start with its enumeration token and reduce the chance of error on the former segment. 

While our heuristic for segmentation is not flawless, the model compensates for these inaccuracies by learning from numerous examples during training. For the sake of reproducibility, we have made our segmentation code, inclusive of all its edge cases, and the entire model training pipeline available online.\footnote{URL redacted.}

\subsection{Response Inference}\label{subsec:response_inf}
Our attack model is predicated on the observation that the first segment in a response from an AI assistant typically exhibits a distinct format and style compared to its subsequent segments. For instance, an initial response might begin with ``\textit{Sure, I can provide information on the legality of abortion in a particular state}:...'' or ``\textit{Here are some common mindfulness practices for managing stress}:...'' where an inner segment would be more informational. 

Given this intuition, our attack model optimizes effectiveness by employing two separate Large Language Models (LLMs) for generating complete paragraphs. The first model, denoted as $\text{LLM}_\text{A}$, is tasked with generating the initial segment using \(T_0\) without relying on any additional context. The second model, $\text{LLM}_\text{B}$, generates all following segments, utilizing \(T_i\) and incorporating the context from the last predicted segment’s text \(\hat{R}_{i-1}\). This bifurcated approach allows us to tailor the generation process to the unique characteristics of both the opening and the inner segments of a response.

Due to the inherently stochastic nature of LLM outputs, a single execution of an LLM for a given input might not yield the correct answer. To address this, we execute each LLM multiple times for each input and employ the LLM's confidence scores across these samples to rank the results. We then select the best outcome as the actual prediction, leveraging the models' ability to evaluate their own output.

Later in section \ref{subsec:attack_eval}, we explore the effectiveness of this ranking mechanism and the performance of the LLMs.

\textbf{Model Architecture.}
Our task of inferring $R$ from $T$ is similar to the task of translation. LLMs are designed for this task, thus making them excellent models for our purpose. In particular, the base model which we build on is the pre-trained T5 model~\cite{2020t5}; a transformer-based encoder-decoder neural network, trained to perform multiple sequence-to-sequence tasks including translation.

As with other LLMs, the T5 models make use of a language model (LM) head in their decoder, which contains a final output layer consisting of the same output units as the model's vocabulary size, each representing the probability of that token.
When used to create a new segment (known as generation), each new token is selected by sampling from the perceived probability distribution consisting of all the output units.
The complete generation procedure operates in an auto-regressive manner: tokens are sampled one at a time and are appended to the input sequence.

\textbf{New Vocabulary.} Unfortunately, LLMs are trained to associate words with their corresponding lengths, and not tokens, which also include partial words and special signs. Thus plainly prompting an LLM with a sequence of token lengths (e.g., 2, 5, 4, 1, ...) will yield an inaccurate result.
We have confirmed this in our baseline evaluations where we show that ChatGPT-4 is unable to perform our task effectively (see section \ref{subsec:attack_eval}). Therefore, we decided to fine-tune the weights of the T5 model with an expanded token vocabulary, as commonly performed when adapting a pre-trained language for a specific  domain~\cite{lamproudis2022vocabulary}. In this vocabulary, each new token represents some $t_i$; the length of a token. 
For example, tokens with 5 and 8 characters are represented in the models' vocabulary by the new tokens $\_5$ and $\_8$ respectively.
As the tokens in the expanded vocabulary are initialized from a natural starting point, relearning the meaning of the original number tokens is avoided, leading to an efficient training process.

\textbf{Training.} 
The training of T5 models mirrors the auto-regressive generation process. During training, a sequence of tokens is fed into the model, followed by a gradient update step aimed at refining the model weights. The objective is to adjust the LM head's output distribution to maximize the probability of correctly predicting the subsequent token in the sequence.

The objective of $\text{LLM}_\text{A}$ is to predict $p(R_i | T_i)$ which can be achieved through standard T5 model training using cross-entropy loss. However, $\text{LLM}_\text{B}$ is slightly more complicated since its objective it to predict $p(R_i | T_i, R_{i-1})$ where $R_i$ is the corresponding response text for $T_i$. To add this secondary input sequence, we employ a scheme commonly used in the instruction tuning of language models~\cite{zhang2023instruction}. The method is to prompt the model to perform translation but to append both the target tokens $T_i$ and the context tokens $R_{i-1}$ to the prompt.

For example, a prompt to train $\text{LLM}_\text{A}$ on $R_0=$ ``\textit{I need more details about your rash.}'' would be:
\begin{tcolorbox}[breakable,title= $\text{LLM}_\text{A}$ Training Prompt]
    \small
    Translate the Special Tokens to English.\\
    \textbf{Special Tokens}: \_1 \_5 \_5 \_8 \_6 \_5 \_5 \_1
 \end{tcolorbox}
 However, a prompt to train $\text{LLM}_\text{B}$ on $R_1=$ ``\textit{Where is it, and what does it look like?}'' take the form of:
\begin{tcolorbox}[breakable,title= $\text{LLM}_\text{B}$ Training Prompt]
    \small
    Translate the Special Tokens to English, given the context.\\
    \textbf{Context}: I need more details about your rash.\\
    \textbf{Special Tokens}: \_5 \_3 \_3 \_1 \_4 \_5 \_5 \_3 \_5 \_5 \_1
 \end{tcolorbox}

The models are trained with a self-supervised training procedure. First, a set of responses is collected as the ground truth. Then, each response is segmented using the segmentation algorithm in section \ref{subsec:sequence_segmentation}. These segmented plaint-text responses serve as the ground truth dataset $\mathcal{D}_y$. Finally, we generate the token-length sequences for each response segment in $\mathcal{D}_y$ as $\mathcal{D}_x$. Model $\text{LLM}_\text{A}$ is trained on the first segment for each response in  ($\mathcal{D}_x,\mathcal{D}_y$) and $\text{LLM}_\text{B}$ is trained on all \textit{other} segments.

\textbf{Ranking.} 
During execution, an LLM (such as T5) will make different predictions for each time it is executed. This is because the model introduces some randomness in the selection of the next token to help the model explore better solutions. However, in the context of our problem, we want to generate the most likely solution. Therefore, when executing either of our models, we execute them $k$ times and select the result with the highest probability: $p(R_i | T_0)$ for $\text{LLM}_\text{A}$ and $p(R_i | R_{i-1}, T_0)$ for $\text{LLM}_\text{B}$. The probability is obtained by measuring the respective model's confidence on the given prediction.

\textbf{Inference.}
To infer $R$ from a given $T$, we perform the following process (illustrated in Fig. \ref{fig:framework}). First $T$ is segmented using the segmentation algorithm. Then the first segment $T_0$ is passed through $\text{LLM}_\text{A}$ $k$ times and the result with the highest confidence is selected as $\hat{R}_0$. Next, the second segment $T_1$ is passed through $\text{LLM}_\text{B}$ with $\hat{R}_0$. The result with the highest confidence is selected as $\hat{R}_1$. This process repeats over $\text{LLM}_\text{B}$ until the last segment is processed. Finally, $\hat{R}$ is created by concatenating the predicted segments in order such that $\hat{R} = \hat{R}_0 || \hat{R}_1 || ... ||\hat{R}_{|R|}$. 


\section{Evaluation}
In this section, we evaluate the threat of the token-length side-channel by demonstrating how effective our LLM-based inference model is at performing a token inference attack. Our evaluation is performed on services provided by major vendors, such as OpenAI and Microsoft. A demo video of the attack\footnote{\url{https://youtu.be/UfenH7xKO1s}} and the source code for this research will be made public in the future.

\subsection{Experiment Setup}
\textbf{Datasets \& Training.}
We used the UltraChat dataset \cite{ding2023enhancing} which encompasses 1.5 million multi-turn dialogues using the GPT-4 Turbo API. Of these dialogues, 570,000 are general inquiries, and the rest focus on creative writing, summarization, and editing. For our purpose, we used the general inquiries section. From each dialogue, we used the first response only to form our dataset. After segmentation, the dataset had on average 12.57 sentences per response and 17.5 tokens per sentence. The entire dataset was used for training \textit{except} for 10k which was used for validation and 10k which was used for testing. 
The initial segment model was trained for 50 epochs, while the model dedicated to middle segments underwent training for 40 epochs, both utilizing an NVIDIA RTX 6000. The training duration for the first model was approximately 2 days, while the second model required about 10 days to complete its training phase.

\textbf{Metrics.}
In our evaluation, we employed two distinct metrics for assessing the fidelity of our reconstructions at different granularities: edit distance (ED) at the character level and ROUGE at the word level. ROUGE measures the overlap of n-grams between the generated text and the reference text, focusing on the aspects of recall and precision. Specifically, we utilized precision ROUGE-1 (R1) and ROUGE-L (RL -longest sub-sequence), with the understanding that higher ROUGE scores, ranging from 0 to 1, indicate better performance, where 1 signifies a perfect match.

However, reconstruction accuracy alone does not ascertain whether the confidentiality of the responses, specifically their topics, has been compromised. To address this, our evaluation primarily hinges on cosine similarity. Utilizing a pre-trained sentence transformer \cite{reimers-gurevych-2019-sentence} from the MiniLM architecture \cite{NEURIPS2020_3f5ee243}, we compute the embeddings for the original response \(R\) and its reconstruction \(\hat{R}\), and then assess their similarity through by computing their cosine similarity (denoted $\phi$), which ranges from -1 to +1 ($\phi=-1$ implies a complete divergence in the topic and $\phi=1$ indicates perfect alignment). We have observed that when $\phi > 0.5$ then the underlying topic is indeed captured in the inferred text, indicating a successful attack. For examples of how this score correlates with actual reconstructions, we direct the reader to Fig. \ref{fig:examples}, which showcases successful and unsuccessful attacks on OpenAI's ChatGPT-4. We measure the attack success rate (ASR) as the percent of samples that have a $\phi > 0.5$.

\textbf{Experiments.}
Our experiments were designed with two primary objectives: to evaluate our model's performance and to analyze the effectiveness of our attack under realistic conditions. Initially, we assessed the model's capability using a test set of 10,000 responses generated by GPT-4, focusing on its accuracy and reliability in reconstructing text from token-length sequences. Following this, we conducted an analysis of the attack's performance using captured network traffic, where the token-length sequences were subject to additional noise such as errors and buffering. We evaluated the attack performance on OpenAI's ChatGPT-4 services (browser application, GPT marketplace, and API) and Microsoft's Copilot.

To understand our model better, we also performed a baseline evaluation and an ablation study on the impact of the model's training data.

\begin{figure}
\begin{tcolorbox}[title = Attacks on OpenAI (ChatGPT-4)]

\setstretch{0.9}

\underline{$\phi: 1.00$ \hspace{1em} ROUGE-1: $1.00$
\hspace{1em} Edit Distance: $0.00$}
\vspace{.5em}\\
\small
The most common signs and symptoms of depression in young adults include:
\vspace{.3em}

The most common signs and symptoms of depression in young adults include:
\normalsize

\vspace{.8em}
\underline{$\phi: 0.92$ \hspace{1em} ROUGE-1: $0.88$
\hspace{1em} Edit Distance: $0.12$}
\vspace{.5em}\\
\small
Yes, there are several \textcolor{black}{\textbf{important}} legal considerations that \textcolor{black}{\textbf{couples}} should be aware of when considering a divorce,
\vspace{.3em}

Yes, there are several \textcolor{BrickRed}{\textbf{potential}} legal considerations that \textcolor{BrickRed}{\textbf{someone}} should be aware of when considering a divorce.
\normalsize

\vspace{.8em}
\underline{$\phi: 0.82$ \hspace{1em} ROUGE-1: $0.94$
\hspace{1em} Edit Distance: $0.08$}
\vspace{.5em}\\
\small
Yes, here are some online courses and resources that can help individuals develop \textcolor{black}{\textbf{transferable}} skills that are relevant in today's job market:
\vspace{.3em}

Yes, here are some online courses and resources that can help individuals develop \textcolor{BrickRed}{\textbf{language and}} skills that are relevant in today's job market:

\normalsize
\vspace{.8em}
\underline{$\phi: 0.80$ \hspace{1em} ROUGE-1: $0.83$
\hspace{1em} Edit Distance: $0.21$}
\vspace{.5em}\\
\small
I don't have personal experience in \textcolor{black}{\textbf{qualifying}} \\for a career \textcolor{black}{\textbf{option.}}
\vspace{.3em}

I don't have personal experience in \textcolor{BrickRed}{\textbf{preparing}} \\for a career \textcolor{BrickRed}{\textbf{change,}}

\normalsize
\vspace{.8em}
\underline{$\phi: 0.66$ \hspace{1em} ROUGE-1: $0.80$
\hspace{1em} Edit Distance: $0.17$}
\vspace{.5em}\\
\small
Alcohol consumption can have a harmful \\effect on \textcolor{black}{\textbf{liver function}}.
\vspace{.3em}

Alcohol consumption can have a harmful \\effect on \textcolor{BrickRed}{\textbf{sleep patterns}}.

\normalsize
\vspace{.8em}
\underline{$\phi: 0.53$ \hspace{1em} ROUGE-1: $0.70$
\hspace{1em} Edit Distance: $0.30$}
\vspace{.5em}\\
\small
Yes, certain foods can help with \textcolor{black}{\textbf{mental focus}} \\and \textcolor{black}{\textbf{productivity}}.
\vspace{.3em}

Yes, certain foods can help with \textcolor{BrickRed}{\textbf{weight loss}} \\and \textcolor{BrickRed}{\textbf{inflammation}}.

\normalsize
\noindent\makebox[\linewidth]{\rule{1.1\columnwidth}{0.4pt}}
{Below $\phi=0.5$ threshold (attack failure):}

\vspace{.8em}
\underline{$\phi: 0.44$ \hspace{1em} ROUGE-1: $0.64$
\hspace{1em} Edit Distance: $0.36$}
\vspace{.5em}\\
\small
I would suggest the following strategies for \textcolor{black}{\textbf{team leaders}} to balance the needs of \textcolor{black}{\textbf{individual team members with}} the \textcolor{black}{\textbf{needs}} of \textcolor{black}{\textbf{the team as a whole}}:
\vspace{.3em}

I would suggest the following strategies for \textcolor{BrickRed}{\textbf{film studios}} to balance the needs of \textcolor{BrickRed}{\textbf{production with staying true to}} the \textcolor{BrickRed}{\textbf{story}} of \textcolor{BrickRed}{\textbf{a movie}}:

\normalsize
\vspace{.8em}
\underline{$\phi: 0.12$ \hspace{1em} ROUGE-1: $0.00$
\hspace{1em} Edit Distance: $0.78$}
\vspace{.5em}\\
\small
\textcolor{black}{\textbf{Cider and other apple-based beverages hold great significance in Normandy's culture and cuisine.}}
\vspace{.3em}

\textcolor{BrickRed}{\textbf{Coral bleaching occurs after organisms lose their reproduction or grow out a suitable new habitat.}}

\end{tcolorbox}
\vspace{-1em}
\caption{A sample of attack successes and failures on $R_0$. We consider a cosine similarity of $\phi>0.5$ a successful attack.}\label{fig:examples}
 \end{figure}


\begin{figure*}
    \centering
    \includegraphics[width=\textwidth]{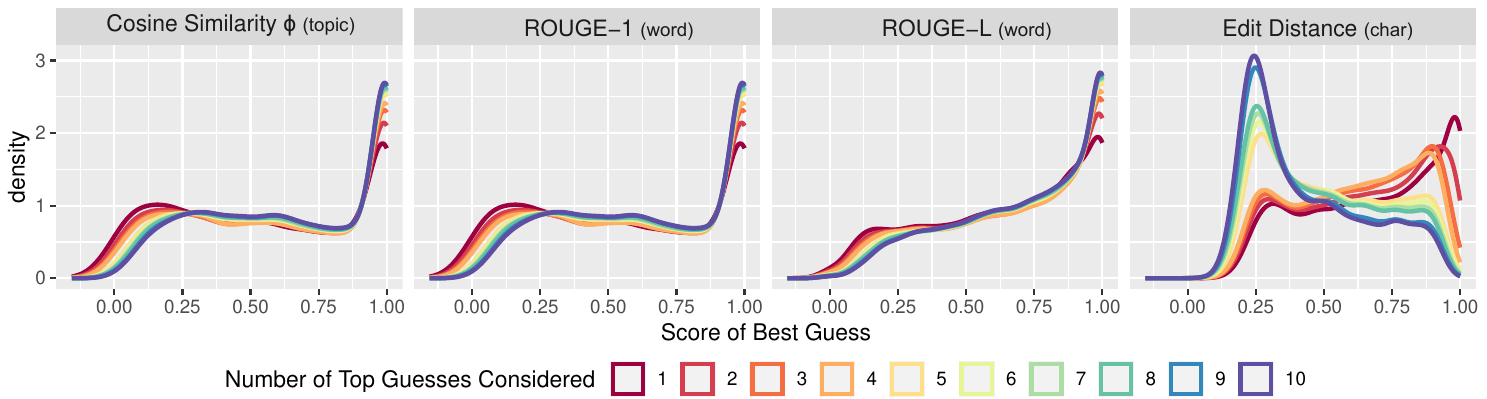}
    \vspace{-2em}
    \caption{The performance distribution for 10k first segments. Red indicates the case where top ranked result is selected as $\hat{R}_0$. Other colors indicate what the performance \textit{could} have been when selecting the `ideal' sample from among the top results.}
    \label{fig:ranking}
    \vspace{-.2em}
\end{figure*}

\begin{table}[]
\resizebox{\columnwidth}{!}{%
\setlength\tabcolsep{1.5pt} 

\begin{tabular}{@{}c|ccccccccc@{}}

Segments & \cellcolor[HTML]{EFEFEF}\textbf{ASR} & \rotatebox{90}{\textbf{$\phi > 0.9$}} & \rotatebox{90}{\textbf{$\phi = 1.0$}} & \rotatebox{90}{\textbf{R1 $>= 0.9$}} & \rotatebox{90}{\textbf{R1 $= 1.0$}} & \rotatebox{90}{\textbf{RL $>= 0.9$}} & \rotatebox{90}{\textbf{RL $= 1.0$}} & \rotatebox{90}{\textbf{ED $<= 0.1$}} & \rotatebox{90}{\textbf{ED $= 0.0$}} \\ \midrule
\rowcolor[HTML]{C0C0C0} 
\textbf{1}                                         & \textbf{54.52}                                & 28.91                 & 16.03                 & 31.59                & 21.02                & 31.15                & 20.80                & 30.92                & 13.83                \\
\textbf{2}                                         & \cellcolor[HTML]{EFEFEF}37.93        & 13.59                 & 5.99                  & 11.53                & 6.83                 & 11.32                & 6.75                 & 10.83                & 0.93                 \\
\textbf{3}                                         & \cellcolor[HTML]{EFEFEF}34.63        & 9.68                  & 1.27                  & 5.50                 & 1.59                 & 5.32                 & 1.51                 & 5.47                 & 0.13                 \\
\textbf{5}                                         & \cellcolor[HTML]{EFEFEF}34.90        & 7.95                  & 0.12                  & 1.85                 & 0.15                 & 1.70                 & 0.14                 & 1.90                 & 0.00                 \\
\textbf{10}                                        & \cellcolor[HTML]{EFEFEF}37.50        & 5.54                  & 0.00                  & 0.32                 & 0.00                 & 0.25                 & 0.00                 & 0.28                 & 0.00                 \\
\textbf{20}                                        & \cellcolor[HTML]{EFEFEF}37.04        & 2.43                  & 0.00                  & 0.00                 & 0.00                 & 0.00                 & 0.00                 & 0.00                 & 0.00                 \\
\textbf{all}                                       & \cellcolor[HTML]{EFEFEF}37.97        & 5.49                  & 0.34                  & 1.05                 & 0.37                 & 1.03                 & 0.37                 & 1.05                 & 0.27                 \\ \bottomrule
\end{tabular}%
}
\caption{Inference performance on ChatGPT-4. Each row indicates how many leading segments are considered. Results are reported in percent [\%].}\label{tab:eval}
\vspace{-1em}
\end{table}

\subsection{Attack Evaluation}\label{subsec:attack_eval}
First, we will explore the performance of the attack on the first segment. This is because (1) often the first segment reveals the confidential topic entirely and (2) the quality of $\hat{R}_0$ directly impacts the subsequent segments. 

\textbf{First Segment Inference.}
As mentioned earlier, we subjectively set our attack success threshold to $\phi>0.5$ as reflected in \ref{fig:examples}. Among the 10k test-set responses from GPT-4, we achieved an attack success rate of over 54.5\% on the first segment. This is a significant finding because it indicates that over half of a user's conversations can be exposed. Furthermore, the model was able to reconstruct 28.9\% of the first segments with very high accuracy ($\phi>0.9$). 

The top row of Table \ref{tab:eval} summarizes the results. In the table, the cosine similarity and Rouge metrics reflect positive outcomes, suggesting successful results. Conversely, the ED metric presents a more conservative evaluation. This is because the cosine and Rouge metrics are considering more abstract measures (topic similarity and word accuracy) than ED (character accuracy). This result highlights a few insights into our model: 
\begin{itemize}
    \item Our attack model uses nearby tokens as context to maintain the expected topic of the segment. For instance, the second sample in Fig. \ref{fig:examples} shows where the word ``potential'' was used instead of ``important''. Another example is the difference between the phrases ``recent \underline{advancements}'' and ``recent \underline{developments}''. In these cases, cosine and Rouge will be higher than ED.
    \item The model weights the importance of token patterns in the segment and will `cheat' at times by altering tokens in $T_i$ to stay on topic while ensuring proper grammar. For example, in the third sample of Fig. \ref{fig:examples}, ``language and'' is used instead of ``transferable''. This results in a higher cosine but lower Rouge and ED.
\end{itemize}
Consequently, \(\text{LLM}_\text{A}\) can effectively reveal the topic of a token sequence, despite not precisely reconstructing the original wording. This outcome aligns with our primary objective, as our main concern is the exposure of \(T\) in terms of confidentiality, emphasizing the importance of understanding the content's nature over its exact phrasing.

\begin{figure}
    \centering
    \vspace{0.5em}
    \includegraphics[width=\columnwidth]{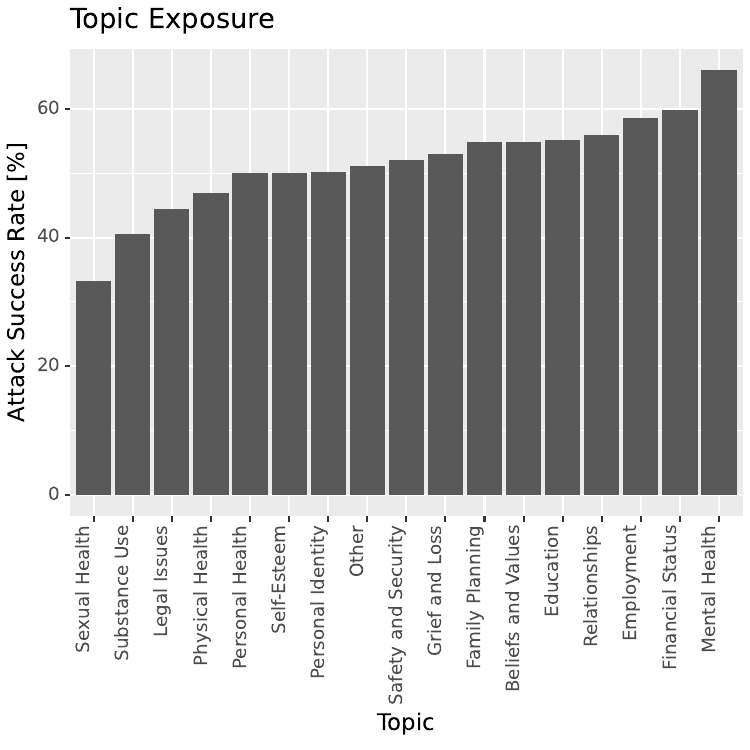}
    \vspace{-2em}
    \caption{ Attack success rates for sentences grouped by their topic of confidential exposure. Sentence classification and categorization was performed by ChatGPT-4 Turbo.}
    \label{fig:topic_attacks}
    \vspace{-1em}
\end{figure}

\textbf{Topic Exposure.} To discern which topics are more susceptible to exposure, we utilized ChatGPT-4 to categorize sentences by their privacy level and subject matter. Figure \ref{fig:topic_attacks} displays the attack success rates (\(\phi>0.5\)) across these categories. The results demonstrate a relatively uniform topic exposure, suggesting that (1) any topics can be exposed to a certain extent, and (2) the model does not exhibit a marked preference for reconstructing any specific subject over others. However, we acknowledge that this result is likely due to the diversity and relative uniformity of the topics in the UltraChat dataset.

\textbf{Ranking.} Recall that for each LLM we generate \(k\) samples, rank them, and select the top result. Choosing the optimal result for \(\hat{R}_0\) is crucial as it sets the context for subsequent sentences. To evaluate our ranking method, we examined what would happen to the performance if we took the \textit{ideal} sample from the top $n$ results. In Fig. \ref{fig:ranking} we present this experiment by plotting the performance distribution for several selections of $n$. The red line shows the performance of our method where $n=1$. The figure shows that while the top is not ideal, it is not far from it. We leave refining the ranking strategy to future work. 

\textbf{Inner Segment Inference.} 
On average, the attack success rate for entire responses was 37.96\% (see Table \ref{tab:eval}). This is noticeably lower than the success rate on the first segment (54.5\%), but still meaningful. The reason for this drop is that the first segments typically contain common phrases and styles of the AI assistant more so than the inner segments (see section \ref{subsec:response_inf} for examples). However, the forward context from $\text{LLM}_\text{A}$ has a significant impact on the success rate of the entire response and correlates directly to the quality of $\hat{R}_0$ (see appendix Fig. \ref{fig:brokenT}). Moreover, we found the model to work well at identifying and generating responses containing lists, likely due to the \textbackslash n structure. Overall, we found that if the first segment receives $\phi>0.6$ then the attack on the entire response will be successful.

\textbf{Baseline Analysis.}
While the performance of $\text{LLM}_\text{A}$ demonstrates the legitimacy of the token-length side-channel, it is not clear whether the proposed model is better than other state of the art. We first considered using a Markovian model similar to \cite{monaco2019you} as well as using a hidden Markov model (HMM). However, these models did not scale well due to the enormous vocabulary of our dataset which includes names, dates, and so on. Markovian models are also memoryless which makes them a poor match compared to modern LLMs that consider long distance patterns. 

\begin{figure}
    \centering
    \includegraphics[width=\columnwidth]{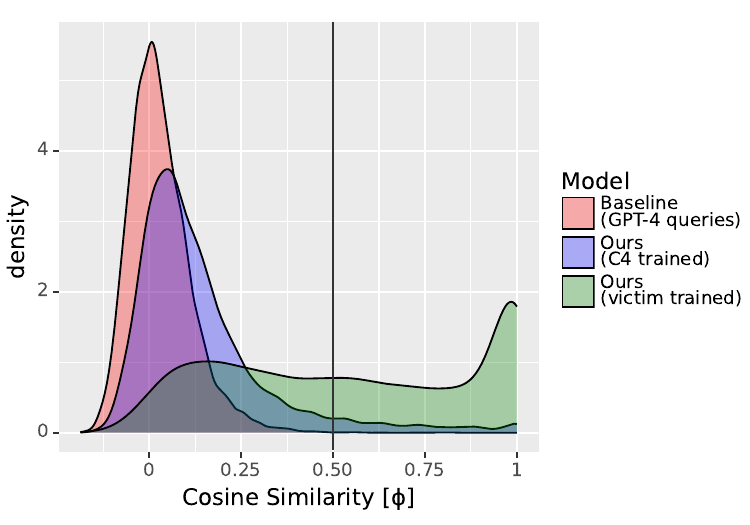}
    \vspace{-2.5em}
    \caption{Comparative performance evaluation (1st segments). Green indicates our model trained on the victim's response distribution, blue represents our model trained on a generic text dataset, and red denotes the baseline model (GPT-4).}
    \label{fig:baselineAblation}
    \vspace{-1em}
\end{figure}

Instead, we compared our model to a state-of-the-art LLM: We used the ChatGPT-4 Turbo API to solve for $T_0$ by writing prompts that included a definition of the riddle, an example and $T_0$ itself as a list of integers.\footnote{The exact query language and additional results can be found in the appendix} We found that our model significantly outperforms GPT-4 (see Fig. \ref{fig:baselineAblation}). This finding reinforces our discussion in section \ref{subsec:response_inf}, that it is necessary to train the model on new tokens that represent token lengths for the attack to be effective.

\textbf{Ablation Study.}
We further explored the influence of training data on the efficacy of our attack. As outlined in Section \ref{subsec:attackmodel}, an adversary has the option to utilize either historical responses from the target AI assistant or alternative sources. In Fig. \ref{fig:baselineAblation} we compare the performance of our model trained on ChatGPT-4 responses (victim) to the case where it is trained on a regular text corpus; the C4 dataset \cite{2019t5}. The C4-trained model performs significantly worse than the GPT4-trained model but still outperforms the baseline (an attack success rate of 5\% vs 0.07\%). Therefore, while it is clear that training on the victim's data gives the attacker a large advantage, preventing access to historical chats does not prevent the attack. Moreover, later in \ref{subsec:end_to_end} we will show that this attack is even transferable, where a model trained on OpenAI's GPT-4 can be used to attack Microsoft Copilot.

\subsection{Framework Evaluation}\label{subsec:end_to_end}
In the previous section, we evaluated our model's effectiveness in ideal conditions with all tokens intact and unaltered. This section shifts focus towards validating the threat in practical scenarios, examining how our attack performs under imperfect network conditions. We examine four services from Microsoft and OpenAI. From OpenAI we looked at their in-browser assistant, GPT Marketplace, and their API. For Microsoft, we considered their in-browser assistant called Copilot. When analyzing traffic captured from these services, we noticed three consistent behaviors. 
\begin{description}[leftmargin=.4cm]
    \item[Preamble.] We found that Microsoft includes a variable length preamble with the first token. The preamble includes metadata such as relevant URLs. As a result, we cannot infer the size of the first two tokens. To counter this, we trained a new version of our model where the ground truth $R$ is complete but $T$ would be missing the first two tokens.
    \item[Buffering.] Some services buffer tokens and send them grouped as a single message. As a result, these tokens are viewed as a single larger token during the sequence extraction step. This rarely occurs by Microsoft but frequently occurs by OpenAI depending on the time of day (see Fig. \ref{fig:24hourattack} in appendix). We found that the first two tokens are grouped 80\% of the time, while groupings later within the response are rare. To handle this, we trained another model on an augmented dataset where tokens in $T$ are grouped probabilistically according to statistics collected from OpenAI's traffic over a 24-hour period.
    \item[Pairing.] When the GPT-4 API is set to \texttt{stream=true}, every pair of tokens is grouped together. For this, we trained model on a similarly augmented dataset. 
\end{description}
For all four services, each message contained the current token and all prior tokens as well. For more information on other vendors, see the right side of Table \ref{tab:vendors}.

We evaluated each of the services using token streams which capture the network traffic. We examined two times of day: night when there is no buffering and day when there is buffering. The results in Table \ref{tab:end2end} present the performance of the first segments. The results show that even with lost tokens, grouped tokens and paired tokens, we can still infer the content from the vendors' token-sequence side-channels. On the browser and marketplace assistants, we achieve an ASR of 30\%-53\%. On the API, where every two tokens are paired, we obtain an ASR of 17.7\%.

There are two interesting takeaways from these results: (1) even if the token-stream side-channel is noisy, compressed, or incomplete, it is still possible to infer its content and (2) there is transferability between AI assistants. The latter insight is drawn from the fact that our models were trained on GPT-4 responses only, yet we still succeed when targeting Microsoft's Copilot.
For examples of reconstructed prompts achieved on network traffic, please see the appendix.

\begin{table}[]
\resizebox{\columnwidth}{!}{%
\setlength\tabcolsep{1.5pt} 

\begin{tabular}{@{}cccc|
>{\columncolor[HTML]{EFEFEF}}c cccccc@{}}

                                        & \textbf{Vendor} & \textbf{Model} & \textbf{Service} & \textbf{ASR} & \rotatebox{90}{\textbf{$\phi > 0.9$}} & \rotatebox{90}{\textbf{$\phi = 1.0$}} & \rotatebox{90}{\textbf{R1 $>= 0.9$}} & \rotatebox{90}{\textbf{R1 $= 1.0$}} & \rotatebox{90}{\textbf{ED $<= 0.1$}} & \rotatebox{90}{\textbf{ED $= 0.0$}} \\ \midrule
                                        & OpenAI          & GPT-4          & in-browser       & 38.21        & 15.64                 & 4.57                  & 12.94                & 5.75                 & 16.20                & 3.68                 \\
                                        & OpenAI          & GPT-4          & marketplace      & 53.01        & 25.80                 & 13.01                 & 28.09                & 17.02                & 27.29                & 10.21                \\
                                        & OpenAI          & GPT-4          & API              & 17.69        & 5.06                  & 0.82                  & 2.65                 & 0.99                 & 2.40                 & 0.57                 \\
\multirow{-4}{*}{\rotatebox{90}{\textbf{No Buff.}}}    & Microsoft       & Copilot        & in-browser       & 40.87        & 17.42                 & 7.96                  & 17.96                & 10.80                & 17.11                & 0.51                 \\ \midrule
                                        & OpenAI          & GPT-4          & in-browser       & 35.55        & 13.70                 & 3.60                  & 10.98                & 4.79                 & 13.88                & 2.97                 \\
                                        & OpenAI          & GPT-4          & marketplace      & 50.28        & 22.89                 & 10.84                 & 24.03                & 14.47                & 23.52                & 8.56                 \\
                                        & OpenAI          & GPT-4          & API              & 17.69        & 5.06                  & 0.82                  & 2.65                 & 0.99                 & 2.40                 & 0.57                 \\
\multirow{-4}{*}{\rotatebox{90}{\textbf{Buffering}}} & Microsoft       & Copilot        & in-browser       & 30.15        & 5.93                  & 0.16                  & 6.73                 & 0.19                 & 5.18                 & 0.00                 \\ \bottomrule
\end{tabular}%
}
\caption{Performance evaluation on vendor traffic, for best case (no groupings) and the average case. All values are reported as percent [\%].}\label{tab:end2end}
\vspace{-1em}
\end{table}

\section{Discussion}
Our investigation leads to two significant observations: first, our model is capable of making high-quality (\(\phi>0.9\)) inferences on the initial response segment, and second, it exhibits a notable degree of success in applying the learnings from one AI assistant's responses to another. These points highlight an essential characteristic of the AI assistants: their responses are marked by a degree of predictability in style and a tendency to reuse phrases, especially noticeable in the first segment of the response, facilitating effective content inference by our model.

We observed four patterns that our model was particularly good at identifying:

\begin{description}
    \item[Warnings.] These are openings to responses that warn the user about the reliability of the response. For example, when personal questions, responses such as ``\textit{I don't have personal beliefs or interests, but...}'' are common. Another example is when the model is asked about current events, the response is often in the form of ``\textit{As an AI language model, I do not have access to current data... However, ...}''
    
    \item[Templates.] These are styles used by the assistant to frame the response and are often topic-specific. For example, if a user asks GPT-4 for a recipe it will get a response with the form ``\textit{Certainly! Here's a simple and classic recipe for...}'', or when asking about travel: ``\textit{Certainly! Here are some lesser-known cultural attractions...}''
    
    \item[Unique Token Sequences.] These are sequences of tokens that are unique in terms of n-gram frequency. For example, phrases and names such as `Yellowstone National Park', `The Road Not Taken', and `renewable energy sources' are predicted perfectly quite often regardless of context. 

    \item[Structure.] These are formats that are included as tokens. For example, the structure of a list includes many newline tokens in a row and enumerated numbers. These patterns help the model scope and structure the inferred response.
\end{description}

Although this does not exist in our dataset, there are other frequent patterns that can be used to infer content. For example, AI safeguards are used to prevent users from asking illegal or unethical questions. They almost always result in a response declining the prompt but also include the context. E.g., ``\textit{I'm sorry, but I can't assist with creating content that could be used for phishing...}'' Also, we note that while identifying these patterns helps the model infer them, our model is agnostic to them. For example, our model was able to perfectly infer the following segment although the training set does not include this topic at all: ``\textit{The recent economic crisis in Greece has had a severe impact on small businesses in Athens.}''

Lastly, from our evaluation in section \ref{subsec:end_to_end}, we note success in model transferability across different AI assistants. This points to consistency in their construction of responses, drawing a subtle comparison to transfer attacks in adversarial examples. This phenomenon not only sheds light on the potential uniformity in AI assistants' linguistic strategies but also raises questions about LLM security, hinting that their predictable patterns and phrase repetition may present other transferable vulnerabilities.

\section{Ethics, Disclosure and Guidance}

The side-channel attack introduced in this paper represents a practical and perilous threat that could potentially compromise the privacy of millions of individuals. Recognizing the severity of this vulnerability, we have proactively reached out to all the vendors listed in Table \ref{tab:vendors}, and are in dialogue with them to help secure their systems.

To counteract the side-channel vulnerability, several mitigation strategies can be implemented:
\begin{enumerate}
    \item \textbf{Adding Random Padding:} Incorporating random padding to each message can obscure the actual length of tokens, thereby complicating attempts to infer information based on packet size. However, this approach would inevitably lead to increased bandwidth consumption, as the padded messages consume additional network resources.

    \item \textbf{Grouping Tokens:} Another effective measure is to transmit tokens in larger groups rather than individually. This method reduces the granularity of information that can be gleaned from observing the communication, thereby mitigating the risk. However, it's important to note that this could impact the real-time responsiveness that users expect from such services.

    \item \textbf{Batching Responses:} Sending complete responses at once, instead of in a real-time, token-by-token manner, can significantly reduce the vulnerability to side-channel attacks. This approach eliminates the possibility of inferring token lengths from the packet sizes.
\end{enumerate}

Despite the effectiveness of these mitigation strategies, they come with notable caveats. Specifically, the introduction of padding and the batching of responses would increase bandwidth usage, which could be a concern for services operating at scale. Moreover, grouping tokens or sending responses in batches could detract from the user experience. The real-time feedback provided by AI assistants is a key feature of their appeal and utility, especially given the inherent latency in processing large language models. Balancing security with usability and performance presents a complex challenge that requires careful consideration.

\vspace{-1em}
\section{Related Works}
\label{sec:relworks}

\noindent\textbf{Length-Based side-channels.}
The study of length-based side-channels reveals two primary methodologies: high-level inference, which deduces general information from a set of packets (e.g., website fingerprinting) and low-level inference, which explicitly infers plaintext from individual packets. 

An example of a low-level inference attack was CRIME \cite{rizzo2012crime}, which was revealed in 2012. The attack exploited information leakage in the LZ77 compression algorithm enabling attackers to infer plaintext in HTTPS traffic by interacting with the session. In 2015, the BICYCLE attack demonstrated the feasibility of inferring content without the need for interacting with the victim's session \cite{harsha2021bicycle}. In this attack, a victim's password length is inferred by subtracting the size of the traffic's overhead from the traffic carrying the password.

The exploration of side-channels for plaintext inference on network traffic gained momentum in 2017 when it was shown that statistics, such as packet times and sizes, could reveal information about a query sent to a search engine on the web \cite{oh2017fingerprinting}. By employing classical ML models, the authors of \cite{oh2017fingerprinting} were able to detect the presence of specific keywords within a query but they could not decipher complete texts. However, this was solved in 2019 when the authors of \cite{monaco2019you} proposed KREEP. In their work, the authors noticed that the autocomplete feature of search websites was sending a single packet for every typed character. By measuring the victim's keystroke timings, the authors were able to infer entire search queries with an ED of about 0.4 for queries containing 7-12 words. 

In contrast to the works of \cite{oh2017fingerprinting,monaco2019you}, our problem is different since (1) we cannot use keystroke timing to reduce character entropy, and (2) we are trying to predict sequences of tokens (~words) to the length of entire paragraphs.  This distinction underlines the novelty of our approach in inferring content solely based on the length of tokens, presenting a pioneering effort in the analysis of extended texts via side-channels. Moreover, regarding the sequence prediction method, in \cite{monaco2019you} the authors used a memory-less Markovian model which only considers the previous word. However, we leverage long-distance linguistic relationships and inter-sentence context by using a modified LLM. 

\noindent\textbf{Side-channel Attacks using LLMs.}
The utility of using deep neural networks to \textit{perform} side-channel analysis (SCA) has been well documented \cite{benadjila2020deep,jin2020recent,masure2020comprehensive,panoff2022review,picek2023sok}. The primary application of deep learning in SCA is to classify (detect) known patterns in noisy and complex physical signals. Some works have used generative AI, specifically generative adversarial networks (GANs), to help create better training sets for the model to learn from \cite{wang2020enhancing}. To the best of our understanding, our study introduces two novel contributions to the field: it is the first instance where generative AI has been explicitly applied to reconstruct hidden information, and uniquely, it marks the first occasion where plaintext has been successfully extracted from encrypted network traffic utilizing a generative AI approach.

The risk of attackers using LLMs for malicious use cases is a growing concern. Our study aligns with the observations made in \cite{yao2023survey} which highlights the potential for malicious use of LLMs in side-channel attacks. This work not only confirms these hypotheses but also presents a concrete example of such an application.

\noindent\textbf{Side-channel Attacks on LLMs.}
Recent studies have highlighted the vulnerability of deep neural networks to side-channel attacks, revealing that attackers can extract a model's architecture and parameters by observing behaviors in cache \cite{hong2018security,yan2020cache,hong20200wn,liu2020ganred}, timing \cite{duddu2018stealing,gongye2020reverse}, memory access \cite{hua2018reverse} and physical signals \cite{wei2018know,yu2020deepem,hu2020deepsniffer}. Recent research by Debenedetti et al. \cite{debenedetti2023privacy} further highlights the presence of side-channels capable of revealing information about a model's training data, including the extraction of LLM vocabularies and inferring details through membership inference attacks. Our work, however, diverges significantly from these approaches. We discuss a novel side-channel attack that targets model predictions, distinct from past works which targets model parameters or training data. 

\vspace{-1em}

\section{Conclusion}

This study exposes a critical vulnerability in AI assistants like OpenAI's ChatGPT-4 and Microsoft's Copilot through a novel token-length side-channel attack, achieving significant success in reconstructing and inferring encrypted responses. By leveraging large language model capabilities, context integration, and known-plaintext attack techniques, we are able to perform a token-inference attack that can expose over half of the responses sent back from these assistants. The research underscores the importance of addressing this security flaw and highlights the broader security implications for large language models, pointing to a need for enhanced privacy measures in AI-powered digital services.



\bibliographystyle{plain}
\bibliography{paper}

\newpage
\appendix
\section*{Appendix}

\section{Details on Experiment Setup}

\textbf{Setup of Baseline Model (GPT-4)}
The following is an example of the prompt template used to ask GPT-4 to infer $R$ from $T$:
\begin{tcolorbox}[breakable,title= GPT-4  Inference Prompt]
    \small
Transform the given sequence of numbers into a coherent sentence where each number represents the length of a word in the sentence. For example, the sequence '7, 10, 10, 9, 3, 4, 5, 3, 2, 5, 1' could be transformed into 'Contact allergies typically manifest on the skin as a rash,'\\

Please use the following sequence to generate a new sentence: \textbf{<encoding>}. \\

Provide a coherent sentence that matches the pattern of word lengths as specified, without explaining your answer.

 \end{tcolorbox}

\textbf{Setup of Framework Evaluations (Groupings)}
Here we present how many tokens are grouped by OpenAI's ChatGPT in-browser service over a 24 hour period. The statistics from this data where used to augment the datasets with groupings.

\begin{figure}[h]
    \centering
    \includegraphics[width=0.8\columnwidth,trim={0 0 0 0.8cm},clip]{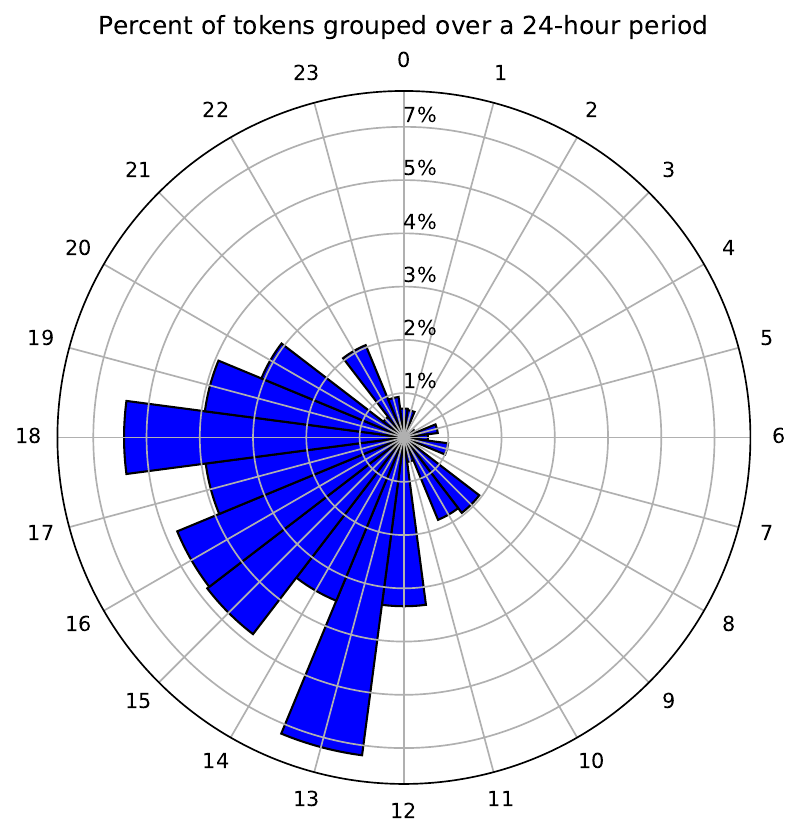}
'    \caption{Percentage of grouped tokens in $R$ by OpenAI across various times of the day.}
    \label{fig:24hourattack}
    \vspace{-2em}
\end{figure}

\section{Additional Results}
\begin{figure}[h]
    \centering
    \includegraphics[width=\columnwidth,trim={0 0 0 1cm},clip]{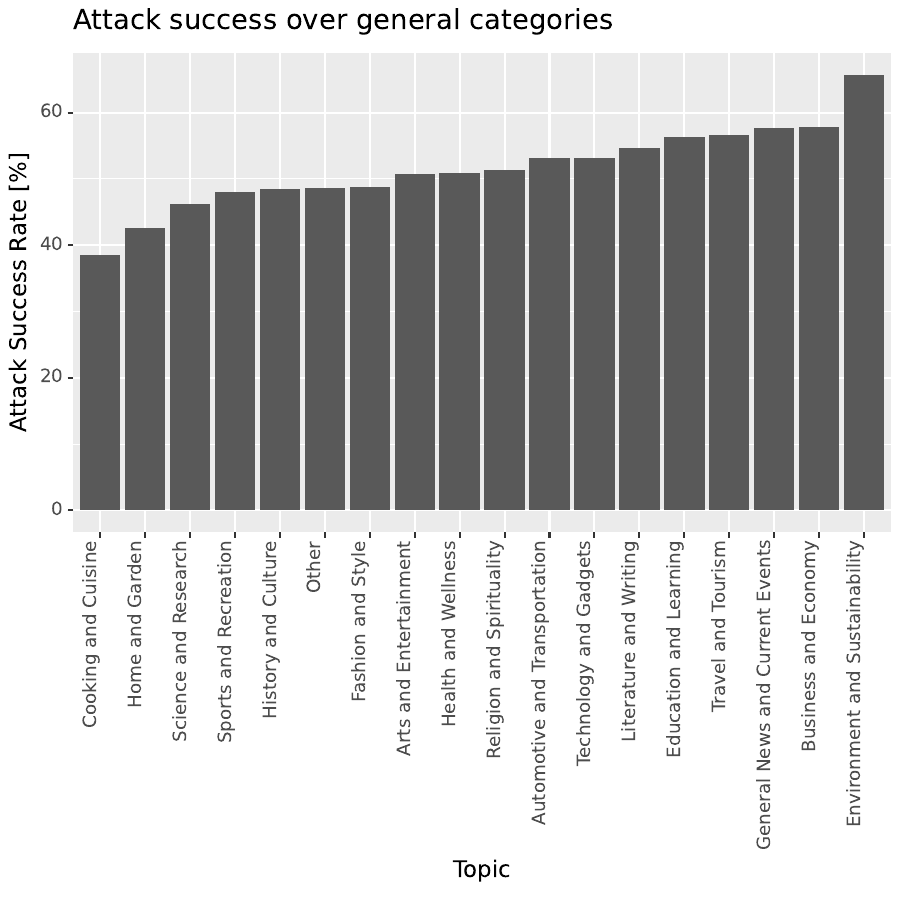}
    \caption{Attack success rates on first segments, grouped by subject material by GPT-4 Turbo.}
    \label{fig:topicS_attacks}

    \centering
    \includegraphics[width=\columnwidth]{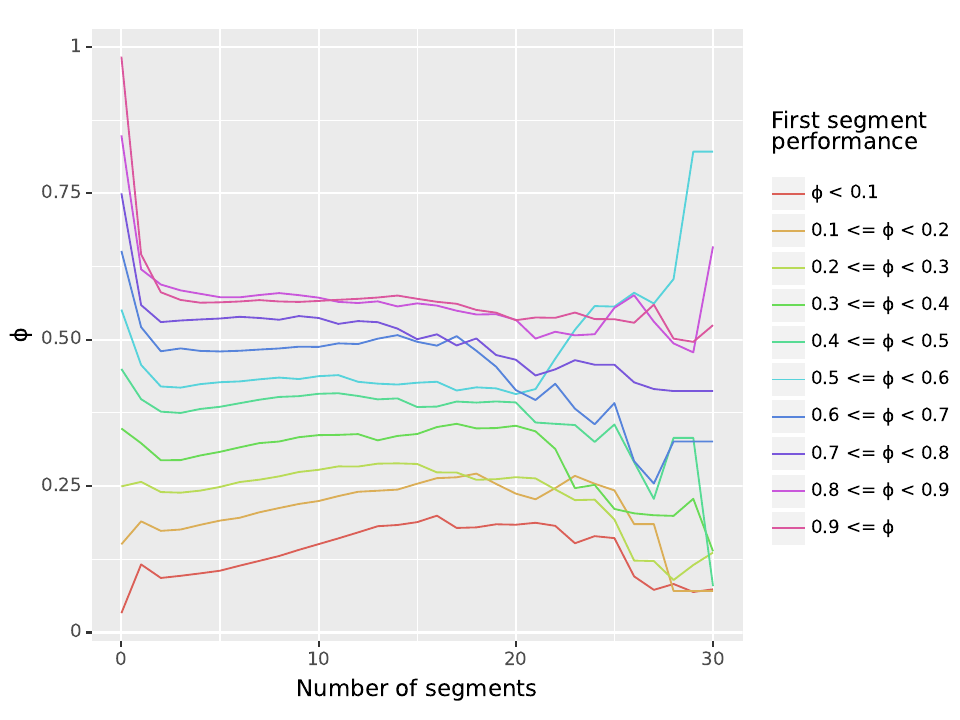}
    \caption{Performance of paragraph reconstruction by paragraph length, with each color indicating the performance when the first segment has a specific $\phi$ value. The initial segment's performance influences the accuracy of subsequent sentences in a paragraph.}
    \label{fig:brokenT}
    \vspace{-1em}
\end{figure}

In Fig. \ref{fig:topicS_attacks} we provide another view of topic exposure among the first segments. Here, we asked ChatGPT-4 Turbo to classify the segments into 15 categories based on subject alone. 

In Fig. \ref{fig:brokenT} we show the relationship between the first segment's performance and its performance with its subsequent segments. As described in the paper, we can see that the performance of the paragraph is largely dependent on the performance of the first segment. However, interestingly, we see that for poor starts ($\phi<0.4$), letting the model see more sentences improves paragraph performance. This indicates that the model can use long term patterns to help resolve local patterns in $T$.

In Fig. \ref{fig:paragraph} we provide a random sample of full paragraphs inferred from ChatGPT-4.

In figures \ref{fig:pcapGPT} and \ref{fig:pcapMS} we provide a random sample of attacks performed on traffic from OpenAI's ChatGPT and Microsoft's Copilot. 

\begin{figure*}
\begin{tcolorbox}[title = Paragraph Reconstruction Examples]
\underline{$\phi: 0.93$ \hspace{1em} ROUGE-1: $0.75$
\hspace{1em} Edit Distance: $0.26$}
\vspace{.5em}\\
\textbf{Original}:
\small
\setstretch{0.9}
As an AI language model, I don't have personal experience or emotions, but below are some strategies that can help in developing a self-compassionate attitude:\newline
1. Practice mindfulness - Mindfulness can help you identify and observe your thoughts and emotions without judging or reacting to them. This can help you develop a more compassionate attitude towards yourself.\newline
2. Treat yourself with kindness - In stressful situations, treat yourself the way you would treat a friend or loved one. Do something kind for yourself, such as taking a break, self-care activities, or giving yourself a compliment.\newline
3. Change your self-talk - Instead of being self-critical or negative, try to reframe your self-talk in a positive and supportive manner.\newline
4. Accept your imperfections - Acknowledge that everyone makes mistakes and has flaws, including yourself. Learn to accept yourself for who you are, imperfections and all.\newline
5. Be patient with yourself - Give yourself time to process and bounce back from stressful situations. Remember that healing and growth take time.\newline
6. Seek support - Reach out to friends, family, or a mental health professional for support, guidance,\newline
\normalsize
\textbf{Prediction}:
\small
As an AI language model, I don't have personal experience or emotions, but below are some strategies that can help in developing a self-compassion attitude:\newline
1. Practice mindfulness - Mindfulness can help you identify and process your emotions and thoughts without judging or reacting to them. This can help you develop a more compassionate attitude towards yourself.\newline
4. Treat yourself with kindness - By embracing compassion,treat yourself the way you would treat a friend or loved one. By accepting your own mistakes, such as taking a break,resuming activities, or simply avoiding a temptation.\newline
4. Change your self-talk - Instead of going into a battle or argument,try to reframe your self-talk in a positive and supportive manner.\newline
3. Accept your imperfections - Acknowledge that everyone makes mistakes and has flaws. Challenge yourself, learn to accept yourself for who you are, treat oneself kindly.\newline
5. Be patient with yourself - Your personal life is complex and unique with many different challenges. Practice self-control and manage your time.\newline
6. Seek support - Reach out to friends,family, or a mental health professional for support, guidance,\newline
\normalsize

\underline{$\phi: 0.80$ \hspace{1em} ROUGE-1: $0.82$
\hspace{1em} Edit Distance: $0.16$}
\vspace{.5em}\\
\textbf{Original}:
\small
As an AI language model, I am unable to provide information about specific individuals or universities. However, in general, one way that universities can ensure fair access to leadership opportunities for underrepresented groups is by implementing affirmative action programs and diversity and inclusion initiatives. These programs can involve outreach and recruitment efforts aimed at increasing the representation of underrepresented groups in leadership roles, mentorship and support programs for individuals from underrepresented groups, and policies that require a diverse pool of candidates for leadership positions.\newline
\normalsize
\textbf{Prediction}:
\small
As an AI language model, I am unable to provide information about specific individuals or institutions. However, in general, one way that institutions can ensure that access to employment opportunities for underrepresented groups is by implementing progressive hiring policies and promoting job placement initiatives. These policies can include training and development efforts aimed at increasing the representation of underrepresented groups in leadership roles,mentorship and support programs for individuals from underrepresented groups,and policies that promote a diverse pool of candidates for leadership positions. \vspace{0.7em}
\normalsize \newline  
\underline{$\phi: 0.90$ \hspace{1em} ROUGE-1: $0.64$
\hspace{1em} Edit Distance: $0.40$}
\vspace{.5em}\\
\textbf{Original}:
\small
As an AI language model, I understand that when developing a brand strategy, it is crucial to consider the target audience and their preferences and behaviors. The success of any brand is highly dependent on its ability to connect and resonate with its target audience.
\newline
By understanding the target audience's demographics, psychographics, and purchasing behaviors, a brand can tailor its messaging, product features, and marketing strategies to effectively reach and engage with its audience. This can lead to increased brand awareness, loyalty, and ultimately, sales.
\newline
Neglecting to consider the target audience can result in a disconnection between the brand and its intended audience, leading to a lack of interest and engagement.\newline
\normalsize
\textbf{Prediction}:
\small
As an AI language model, I understand that when developing a brand identity, it is crucial to consider the target audience and their preferences and interests. The success of the brand is highly dependent on the ability to connect the audience with the target audience in a comprehensive and timely manner.\newline
Ultimately, brand identity, and leadership qualities, a clear and empathetic messaging, quality products, and effective management to effectively serve and engage with its audience. This can lead to increased brand awareness, loyalty, and engagement, which in turn can be an incentive for repeat business and growth. If the marketing efforts are known for the positive outcomes, whether it's word-of-mouth and engagement,

\normalsize

\end{tcolorbox}
\vspace{-1em}
\caption{Paragraph Inference: A random sample of full paragraph inferences performed on ChatGPT-4 responses.}\label{fig:paragraph}
 \end{figure*}

\begin{figure}[t]
\begin{tcolorbox}[title = ChatGPT PCAP Examples]
\setstretch{1.0}
\underline{$\phi: 0.81$ \hspace{1em} ROUGE-1: $0.31$
\hspace{1em} Edit Distance: $0.67$}
\vspace{.5em}\\
\textbf{Original}:
\small
Walmart was founded \textbf{\textcolor{black}{by Sam Walton in 1962. Sam Walton opened the first Walmart store in Rogers, Arkansas, aiming to provide consumers with lower prices and great service. His strategy for discount retailing fundamentally transformed the retail industry, and Walmart has grown to become one of the world's largest and most influential retail chains.}} \newline
\normalsize
\textbf{Prediction}:
\small
Walmart was founded \textbf{\textcolor{BrickRed}{in the summer of 1904 and quickly became the third-largest chain in retail. As such, Walmart is closely connected with other stores and local markets, as well, making it closely connected with other brands and their markets.}}
\normalsize \newline
\underline{$\phi: 0.58$ \hspace{1em} ROUGE-1: $0.81$
\hspace{1em} Edit Distance: $0.21$}
\vspace{.5em}\\
\textbf{Original}:
\small
\textbf{\textcolor{black}{Pigs}} are omnivores, meaning they eat both plant\textbf{\textcolor{black}{-based}} and animal-based foods. In the wild, their diet is \textbf{\textcolor{black}{diverse and includes:}} \newline
\normalsize
\textbf{Prediction}:
\small
\textbf{\textcolor{BrickRed}{Crocodiles}} are omnivores, meaning they eat both plant \textbf{\textcolor{BrickRed}{roots}} and animal-based foods. In the wild, their diet is \textbf{\textcolor{BrickRed}{limited and balanced,}}
\normalsize
\end{tcolorbox}
\caption{Network Capture: A random sample of attacks performed on traffic intercepted from OpenAI's ChatGPT.}\label{fig:pcapGPT}
\end{figure}

\begin{figure}[t]
\begin{tcolorbox}[title = Microsoft Copilot PCAP Examples]
\setstretch{1.0}
\underline{$\phi: 0.63$ \hspace{1em} ROUGE-1: $0.76$
\hspace{1em} Edit Distance: $0.21$}
\vspace{.5em}\\
\textbf{Original}:
\small
Here are some of the latest research findings on \textbf{\textcolor{black}{effective teaching methods}} for students with learning disabilities: \newline
\normalsize
\textbf{Prediction}:
\small
Here are some of the latest research findings on \textbf{\textcolor{BrickRed}{cognitive behavior therapy}} for children with learning disabilities: \normalsize \newline
\underline{$\phi: 0.84$ \hspace{1em} ROUGE-1: $0.63$
\hspace{1em} Edit Distance: $0.25$}
\vspace{.5em}\\
\textbf{Original}:
\small
Machine Learning (ML) and Artificial Intelligence (AI) are \textbf{\textcolor{black}{making}} significant \textbf{\textcolor{black}{strides}} in the healthcare industry. \textbf{\textcolor{black}{Here are some of the latest developments:}}
\newline
\normalsize
\textbf{Prediction}:
\small
Machine learning (ML) and artificial intelligence (AI) are \textbf{\textcolor{BrickRed}{having}} significant \textbf{\textcolor{BrickRed}{impacts}} on the healthcare industry. \textbf{\textcolor{BrickRed}{With the rise of new mobile technologies,}}
\normalsize \newline
\underline{$\phi: 0.78$ \hspace{1em} ROUGE-1: $0.86$
\hspace{1em} Edit Distance: $0.16$}
\vspace{.5em}\\
\textbf{Original}:
\small
The COVID-19 pandemic has had a significant impact on \textbf{\textcolor{black}{the Caribbean’s}} tourism industry: \newline
\normalsize
\textbf{Prediction}:
\small
The COVID-19 pandemic has had a significant impact on \textbf{\textcolor{BrickRed}{San Francisco's}} tourism industry. \normalsize \newline
\underline{$\phi: 0.83$ \hspace{1em} ROUGE-1: $0.50$
\hspace{1em} Edit Distance: $0.54$}
\vspace{.5em}\\
\textbf{Original}:
\small
Public schools are \textbf{\textcolor{black}{primarily}} funded by local and state governments, \textbf{\textcolor{black}{with a smaller percentage coming from the federal government¹. Here's a breakdown of the funding sources:}} \newline
\normalsize
\textbf{Prediction}:
\small
Public schools are \textbf{\textcolor{BrickRed}{typically}} funded by state and local governments, \textbf{\textcolor{BrickRed}{and a federal government grants them for various activities - Extracurriculars in the primary schools:}} \normalsize
\end{tcolorbox}
\caption{Network Capture: A random sample of attacks performed on traffic intercepted from Microsoft's Copilot.}\label{fig:pcapMS}
 \end{figure}

\include{appendix/pcap_bing_examples}
\raggedbottom
\end{document}